\documentclass[a4paper,12pt]{article}

\usepackage{jcappub} 
\usepackage{bm}
\usepackage{changes}
\usepackage{amsmath}
\usepackage{graphicx}
\usepackage{adjustbox}
\usepackage{multirow}
\usepackage[symbol]{footmisc}

\title{Constraints on primordial curvature power spectrum with pulsar timing arrays}

\author{Zhi-Qiang You,$^{a, b}$}
\author{Zhu Yi,\note{Corresponding author.}$^{b,*}$}
\author{and You Wu$^{c,d,*}$}
\affiliation{$^a$Henan Academy of Sciences, Zhengzhou 450046, Henan, China}
\affiliation{$^b$Advanced Institute of Natural Sciences, Beijing Normal University, Zhuhai 519087, China}
\affiliation{$^c$College of Mathematics and Physics, Hunan University of Arts and Science, Changde, 415000, China}
\affiliation{$^d$Department of Physics and Synergistic Innovation Center for Quantum Effects and Applications, Hunan Normal University, Changsha, Hunan 410081, China}
\emailAdd{you\_zhiqiang@whu.edu.cn}
\emailAdd{yz@bnu.edu.cn}
\emailAdd{youwuphy@gmail.com}

\abstract{The stochastic signal detected by NANOGrav, PPTA, EPTA, and CPTA can be explained by the scalar-induced gravitational waves.   In order to determine the scalar-induced gravitational waves model that best fits the stochastic signal, we employ both single- and double-peak parameterizations for the power spectrum of the primordial curvature perturbations, where the single-peak scenarios include the $\delta$-function, box, lognormal, and broken power law model, and the double-peak scenario is described by the double lognormal form.  
	Using Bayesian inference, we find that there is no significant evidence for or against the single-peak scenario over the double-peak model, with $\log$ (Bayes factors) among these models  $\ln \mathcal{B} < 1$. 
	Therefore, we cannot distinguish the different shapes of the power spectrum of the primordial curvature perturbation with the current sensitivity of pulsar timing arrays.
	
}

\begin{document}
	\maketitle
	\flushbottom
	
	\section{\label{intro}Introduction}
	Recently, four pulsar timing array (PTA) collaborations, namely NANOGrav~\cite{NANOGrav:2023hde, NANOGrav:2023gor}, PPTA~\cite{Zic:2023gta,Reardon:2023gzh}, EPTA~\cite{Antoniadis:2023lym,Antoniadis:2023ott}, and CPTA~\cite{Xu:2023wog}, all announced the strong evidence of a stochastic signal consistent the Hellings-Downs angular correlations, pointing to the gravitational-waves (GW) origin of this signal. Assuming the signal originates from an ensemble of binary supermassive black hole inspirals and a fiducial $f^{-2/3}$ characteristic-strain spectrum, the strain amplitude is estimated to be at the order of $\sim 10^{-15}$ at a reference frequency of $1\,\rm{yr}^{-1}$~\cite{NANOGrav:2023gor,Reardon:2023gzh,Antoniadis:2023ott,Xu:2023wog}.  However, the origin of this signal, whether from supermassive black hole binaries or other cosmological sources, is still under investigation~\cite{NANOGrav:2023hvm,Antoniadis:2023xlr, Franciolini:2023pbf,Liu:2023ymk,Vagnozzi:2023lwo,Cai:2023dls,Wang:2023ost,Yi:2023mbm,Bi:2023tib,Wu:2023hsa,Zhu:2023faa,Franciolini:2023wjm,Jin:2023wri,Liu:2023pau,Yi:2023npi,Yi:2023tdk,You:2023rmn,Wu:2023rib,Bi:2023ewq,Chen:2023uiz,Chen:2023swcs}. A promising candidate to explain the signal is the scalar-induced gravitational waves (SIGWs) accompanying the formation of primordial black holes~\cite{Zeldovich:1967lct,Hawking:1971ei,Carr:1974nx,Chen:2018czv,Chen:2018rzo,Liu:2018ess,Liu:2019rnx,Chen:2019irf,Liu:2020cds,Liu:2020vsy,Liu:2020bag,Wu:2020drm,Chen:2021nxo,Liu:2022wtq,Chen:2022fda,Chen:2022qvg,Liu:2022iuf,Zheng:2022wqo,Yi:2022ymw}. Other physical phenomena (see e.g. \cite{Zhu:2018lif,Li:2019vlb,Chen:2021wdo,Wu:2021kmd,Chen:2021ncc,Chen:2022azo,PPTA:2022eul,IPTA:2023ero,Wu:2023pbt,Wu:2023dnp,InternationalPulsarTimingArray:2023mzf,Chen:2023zkb}) can also be the sources in the PTA band.
	
	The SIGW is sourced from scalar perturbations generated during the inflationary epoch  \cite{tomita1967non,Saito:2008jc,Young:2014ana,Yuan:2019udt,Yuan:2019wwo,Chen:2019xse,Yuan:2019fwv,Ananda:2006af,Baumann:2007zm,Alabidi:2012ex,Nakama:2016gzw,Kohri:2018awv,Cheng:2018yyr,Cai:2019amo,Cai:2018dig,Cai:2019elf,Cai:2019bmk,Cai:2020fnq,Pi:2020otn,Domenech:2020kqm,Liu:2021jnw,Papanikolaou:2021uhe,Papanikolaou:2022hkg,Meng:2022low}. They offer valuable insights into the physics of the early Universe and can be detected not only by PTAs but also by space-based GW detectors such as LISA~\cite{Danzmann:1997hm,Audley:2017drz}, Taiji \cite{Hu:2017mde}, TianQin \cite{Luo:2015ght, Gong:2021gvw}, and DECIGO~\cite{Kawamura:2011zz}.   
	Significant SIGWs require the amplitude of the power spectrum of the primordial curvature perturbations to be around $\mathcal{A}_{\zeta}\sim \mathcal{O}(0.01)$ which is approximately seven orders of magnitude larger than the constraints from large-scale measurements of cosmic microwave background (CMB) anisotropy observation, $\mathcal{A}_{\zeta}= 2.1\times 10^{-9}$ \cite{Akrami:2018odb}. Therefore, to account for the observed gravitational wave signal detected by PTAs, the curvature power spectrum must possess at least one high peak. This can be achieved through inflation models with a transition ultra-slow-roll phase~\cite{Martin:2012pe, Motohashi:2014ppa,Yi:2017mxs, Garcia-Bellido:2017mdw, Germani:2017bcs, Motohashi:2017kbs,Ezquiaga:2017fvi, Gong:2017qlj, Ballesteros:2018wlw,Dalianis:2018frf, Bezrukov:2017dyv, Kannike:2017bxn, Lin:2020goi,Lin:2021vwc,Gao:2019sbz,Gao:2020tsa,Gao:2021vxb,Yi:2020kmq,Yi:2020cut,Yi:2021lxc,Yi:2022anu,Zhang:2020uek,Pi:2017gih,Kamenshchik:2018sig,Fu:2019ttf,Fu:2019vqc,Dalianis:2019vit,Gundhi:2020zvb,Cheong:2019vzl,Zhang:2021rqs, Zhang:2021vak,Kawai:2021edk,Cai:2021wzd,Chen:2021nio,Zheng:2021vda,Karam:2022nym,Ashoorioon:2019xqc}. 
	
	To characterize a single-peak primordial curvature power spectrum, various parameterizations such as the $\delta$-function form, box form, lognormal form, or broken power law form are employed.  Among them, the $\delta$-function, box and lognormal parameterizations are investigated in Ref. \cite{NANOGrav:2023hvm}, where the constraints from the PTAs data on the parameters of these models are also given.  The constraints on the broken power law form are provided in Ref. \cite{Franciolini:2023pbf}, where the role of non-Gaussianity is also considered. 
	However, the analysis does not determine which model among these is the most compatible with the PTA signal.
	For the multi-peak primordial curvature power spectrum model \cite{Zhang:2021vak}, we  
	parameterize the primordial curvature power spectrum with the double lognormal form.  In this study, we aim to determine whether the PTA signal favors a single-peak or multi-peak primordial curvature power spectrum and identify the most compatible model with the PTA signal. 
	
	The organization of this paper is as follows: section~\ref{SIGW} provides a brief review of the scalar-induced gravitational waves. Section~\ref{model} presents the constraints on the power spectrum for different forms and identifies the best-fitted model based on the PTAs signal. Finally, section~\ref{discussion} summarizes our findings and provides concluding remarks.
	
	\section{\label{SIGW}Scalar-induced gravitational waves} 
	The large scalar perturbations seeded from the primordial curvature perturbation generated during inflation can act as the source to induce GWs at the radiation domination epoch where the equation of state is $w=1/3$.  In this section, we give a brief review of the SIGW.  In the cosmological background,  the  metric with perturbation in   Newtonian gauge is \cite{Malik:2008im}
	\begin{equation}
		d s^2= -a^2(\eta)(1+2\Phi)d\eta^2 +a^2(\eta)\left[(1-2\Phi)\delta_{ij}+\frac12h_{ij}\right]d x^i d x^j,
	\end{equation}
	where $a$ is the scale factor of the Universe, $\eta$ is the conformal time, $d\eta =dt/a(t)$, $\Phi$ is the Bardeen potential,  
	and $h_{ij} $ are the  tensor perturbations.  The tensor perturbations in the Fourier space can be obtained by the transform
	\begin{equation}
		\label{hijkeq1}
		h_{ij}(\bm{x},\eta)=\int\frac{ d^3k  e^{i\bm{k}\cdot\bm{x}}}{(2\pi)^{3/2}} 
		[h_{\bm{k}}(\eta)e_{ij}(\bm{k})+\tilde{h}_{\bm{k}}(\eta)\tilde{e}_{ij}(\bm{k})],
	\end{equation}
	where the plus and cross polarization tensors $e_{ij}(\bm{k})$ and $\tilde{e}_{ij}(\bm{k})$ are 
	\begin{gather}
		e_{ij}(\bm{k})=\frac{1}{\sqrt{2}}\left[e_i(\bm{k})e_j(\bm{k})-\tilde{e}_i(\bm{k})\tilde{e}_j(\bm{k})\right], \\
		\tilde{e}_{ij}(\bm{k})=\frac{1}{\sqrt{2}}\left[e_i(\bm{k})\tilde{e}_j(\bm{k})+\tilde{e}_i(\bm{k})e_j(\bm{k})\right],
	\end{gather}
	and the basis vectors satisfying $\bm e\cdot \tilde{\bm e}=\bm e \cdot \bm{k}= \tilde{\bm e}\cdot\bm{k}$.
	
	With the source from the   second order of   linear scalar perturbations,  the tensor perturbations with either  polarization  in the Fourier space  satisfy \cite{Ananda:2006af, Baumann:2007zm}
	\begin{equation}
		\label{eq:hk}
		h''_{\bm{k}}+2\mathcal{H}h'_{\bm{k}}+k^2h_{\bm{k}}=4S_{\bm{k}},
	\end{equation}
	where $\mathcal{H}=a'/a $ is the conformal Hubble parameter and a  prime denotes the derivative with respect to the conformal time $\eta$. The second order source  $S_{\bm{k}}$ is  \cite{Ananda:2006af,Baumann:2007zm,Kohri:2018awv} 
	\begin{equation}
		\label{hksource}
		S_{\bm{k}}= \int \frac{d^3\tilde{k}}{(2\pi)^{3/2}}e_{ij}(\bm{k})\tilde{k}^i\tilde{k}^j
		\left[2\Phi_{\tilde{\bm{k}}}\Phi_{\bm{k}-\tilde{\bm{k}}} \phantom{\frac{1}{2}}+ 
		\frac{1}{\mathcal{H}^2} \left(\Phi'_{\tilde{\bm{k}}}+\mathcal{H}\Phi_{\tilde{\bm{k}}}\right)
		\left(\Phi'_{\bm{k}-\tilde{\bm{k}}}+\mathcal{H}\Phi_{\bm{k}-\tilde{\bm{k}}}\right)\right].
	\end{equation}
	The  Bardeen potential in the Fourier space, $\Phi_{\bm{k}}$,  can be connected to the primordial curvature perturbations $\zeta_{\bm{k}}$ produced during inflation epoch through the transfer function,
	\begin{equation}
		\Phi_{\bm{k}}=\frac{2}{3}T(k\eta) \zeta_{\bm{k}},
	\end{equation}
	and  the transfer function   $T(k\eta)$ satisfy
	\begin{equation}\label{transfer}
		T(k\eta)=3\left[\frac{\sin\left({k \eta }/{\sqrt{3}}\right)-\left({k\eta}/{\sqrt{3}}\right)
			\cos\left({k\eta}/{\sqrt{3}}\right)}{\left({k\eta}/{\sqrt{3}}\right)^3}\right].
	\end{equation}
	The equation of the tensor perturbations \eqref{eq:hk} can be solved by the Green's function method and the solution is 
	\begin{equation}\label{hk:green}
		h_k(\eta)=\frac{4}{a(\eta)}\int_{\eta_k}^{\eta}d \tilde{\eta}g_k(\eta,\tilde{\eta})a(\tilde{\eta})S_k(\tilde{\eta}),
	\end{equation}
	where $g_k$ is  the   corresponding  Green's function  satisfying the equation  
	\begin{equation}
		g_k''(\eta,\tilde{\eta}) +\left(k^2-\frac{a''}{a}\right)g_k(\eta,\tilde{\eta}) =\delta(\eta-\tilde{\eta}).
	\end{equation}
	
	During the radiation domination, the scale factor satisfies $a\propto \eta$ and $a''/a=0$, the Green's function is
	\begin{equation}\label{green}
		g_k(\eta,\tilde{\eta})=\frac{\sin\left[k(\eta-\tilde{\eta})\right]}{k}.
	\end{equation}
	The  definition of the power spectrum of  tensor perturbations $h_{k}$ is 
	\begin{equation}
		\label{eq:pwrh}
		\langle h_{\bm{k}}(\eta)h_{\tilde{\bm{k}}}(\eta)\rangle
		=\frac{2\pi^2}{k^3}\delta^{(3)}(\bm{k}+\tilde{\bm{k}})\mathcal{P}_h(k,\eta).
	\end{equation}
	Combining it with the solution of   $h_k$, equation \eqref{hk:green}, we have  \cite{Baumann:2007zm,Ananda:2006af,Kohri:2018awv,Espinosa:2018eve,Lu:2019sti}
	\begin{equation}\label{ph}
		\begin{split}
			\mathcal{P}_h(k,\eta)=&
			4\int_{0}^{\infty}dv\int_{|1-v|}^{1+v}du \left[\frac{4v^2-(1-u^2+v^2)^2}{4uv}\right]^2\\ &\times I_{RD}^2(u,v,x)\mathcal{P}_{\zeta}(k v)\mathcal{P}_{\zeta}(ku),
		\end{split}
	\end{equation}
	where $u=|\bm{k}-\tilde{\bm{k}}|/k$, $v=\tilde{k}/k$, $x=k\eta$,  and $\mathcal{P}_\zeta$ is the power spectrum of the curvature perturbation which is parameterized in the following section. 
	The integral kernel $I_{\text{RD}}$  is \cite{Kohri:2018awv}
	\begin{equation}\label{inte:I}
		I_{\text{RD}}(u, v, x) = \int_{0}^{x} d \bar{x} \frac{\bar{x}}{x} \sin(x-\bar{x})f(v,u,\bar{x})
	\end{equation}
	where 
	\begin{equation}
		f(v,u,x) =\frac{4}{3}T(vx)T(u x)+\frac{4}{9}x\partial_x[T(vx)T(ux)] +\frac{4}{9}x^2\partial_x T(vx)\partial_x T(ux).
	\end{equation}
	
	The definition of the energy density of gravitational waves is 
	\begin{equation}
		\label{density}
		\Omega_{\mathrm{GW}}(k,\eta)=\frac{1}{24}\left(\frac{k}{aH}\right)^2\overline{\mathcal{P}_h(k,\eta)}.
	\end{equation}
	By combining the equation \eqref{ph}  and the definition  \eqref{density},  we obtain \cite{Espinosa:2018eve,Lu:2019sti}
	\begin{equation}
		\label{SIGWs:gwres1}
		\begin{split}
			\Omega_{\mathrm{GW}}(k,\eta)=&\frac{1}{6}\left(\frac{k}{aH}\right)^2\int_{0}^{\infty}dv\int_{|1-v|}^{1+v}du\times\left[\frac{4v^2-(1-u^2+v^2)^2}{4uv}\right]^2 \\
			& \times\overline{I_{\text{RD}}^{2}(u, v, x)} \mathcal{P}_{\zeta}(kv)\mathcal{P}_{\zeta}(ku),
		\end{split}
	\end{equation}
	where $\overline{I_{\text{RD}}^{2}}$  represents the oscillation time average of the integral kernel.  After the horizon reentry  
	and the scale  is well within  the horizon, $k/aH\gg1$ and $x\gg1$, the oscillate  average $\overline{I_{\text{RD}}^{2}}$ can expressed as \cite{Kohri:2018awv}
	\begin{align}\label{I:overline}
		\overline{I_{\text{RD}}^2(v,u,x\gg1)} =& \frac{1}{2x^2} \left[ \frac{3(u^2+v^2-3)}{4 u^3 v^3} \right]^2 \bigg[ \left( -4uv+(u^2+v^2-3) \log \left| \frac{3-(u+v)^2}{3-(u-v)^2} \right| \right)^2   \nonumber \\
		&  \qquad \qquad \qquad + \pi^2 (u^2+v^2-3)^2 \Theta ( v+u-\sqrt{3}) \bigg].
	\end{align} 
	In the radiation-dominated era, $k/aH =x$,  the term $(k/aH)^2$ in the front of equation \eqref{SIGWs:gwres1} can be balanced by the $1/x^2$ term in the front of equation \eqref{I:overline}, implying that the GW energy density expression in equation \eqref{SIGWs:gwres1} becomes independent on the conformal time and only dependent on the scale $k$.  
	
	After generation, the energy density of gravitational waves evolves in the same way as radiation. Using this property, it is straightforward to determine the energy density of gravitational waves at present.   The relation of the GW energy density at present and the generation is \cite{Ando:2017veq}
	\begin{equation}\label{omegagw:rel}
		\Omega_{\text{GW}}(k,\eta_0) = \left[\frac{a^2(\eta)H(\eta)}{a^2_0 H_0}\right]^2  \Omega_{\text{GW}}(k,\eta),
	\end{equation}
	where the subscript ``0" denotes the value at present.  Assuming the entropy of the Universe is conserved, we can obtain \cite{Yi:2020cut}
	\begin{equation}\label{ha:relation}
		H^2(\eta)= a^{-4}(\eta)\left(\frac{g_{*,s}^4g_*^{-3}}{g_{*,s0}^4g_{*,0}^{-3}}\right)^{-1/3} H_0^2 \Omega_{r,0},
	\end{equation}
	where $\Omega_{r,0}$ is the current energy density of the radiation, $g_*$ is the number of relativistic degrees of freedom, $g_{*,s}$ is the number of entropy degrees of freedom, $g_{*,0} = 3.36$, $g_{*,s0} = 3.91$, and $a_0=1$. 
	Substituting equation \eqref{ha:relation} into equation \eqref{omegagw:rel}, we obtain the energy density of the SIGWs at present
	\begin{equation}\label{d}
		\Omega_{\mathrm{GW}}(k,\eta_0)= c_g\Omega_{r,0} \Omega_{\mathrm{GW}}(k,\eta),
	\end{equation}
	where \cite{Vaskonen:2020lbd,DeLuca:2020agl}
	\begin{equation}\label{gwcg}
		c_g=0.387\left(\frac{g_{*,s}^4g_*^{-3}}{106.75}\right)^{-1/3}.
	\end{equation}
	
	\section{\label{model}Models and results}
	At large scales, the observational data from the CMB impose a constraint on the amplitude of the primordial curvature power spectrum, which is limited to $\mathcal{A}_\zeta = 2.1 \times 10^{-9}$ \cite{Akrami:2018odb}. However, there are minimal constraints on the primordial curvature power spectrum at small scales. Consequently, in order to generate significant  SIGWs, it is necessary to enhance the primordial curvature power spectrum to approximately $\mathcal{A}_\zeta \sim \mathcal{O}(0.01)$ at small scales. Thus, the profile of the primordial curvature spectrum exhibits at least one pronounced peak at intermediate scales, while displaying lower amplitudes at both large and very small scales.   In this section, we consider the primordial curvature spectrum with single-peak and double-peak, respectively.  For the single peak, the  commonly employed parameterizations of the primordial curvature spectrum  are the simple $\delta$ function form
	\begin{equation}\label{eq:delta}
		\mathcal{P}_\zeta = A\delta(\ln k -\ln k_p),
	\end{equation}
	the box form
	\begin{equation}\label{eq:box}
		\mathcal{P}_\zeta = A \Theta(k - k_{\mathrm{min}})\, \Theta(k_{\mathrm{max}} - k),
	\end{equation}
	the lognormal form
	\begin{equation}\label{eq:ln}
		\mathcal{P}_\zeta = \frac{A}{\sqrt{2\pi}\Delta} \exp\left[-\frac{1}{2}\left(\frac{\ln k -\ln k_p}{\Delta}\right)^2\right],
	\end{equation}
	and the broken power law form
	\begin{equation}\label{eq:bp}
		\mathcal{P}_\zeta =\frac{A(\alpha+\beta)}{\beta(k/k_p)^{-\alpha}+\alpha(k/k_p)^\beta}+A_*(k/k_*)^{n_{s_*}-1},
	\end{equation}
	with $n_{s_*}=0.965$, $k_* = 0.05 \mathrm{Mpc}^{-1}$, and $A_*  =2.1\times 10^{-9}$ \cite{Planck:2018jri}. 
	For the double peak model, we parameterize the primordial curvature spectrum with a double lognormal form 
	\begin{equation}\label{eq:dln}
		\mathcal{P}_\zeta= \frac{A_1}{\sqrt{2\pi}\Delta_1} \exp\left[-\frac{1}{2}\left(\frac{\ln k -\ln k_{p_1}}{\Delta_1}\right)^2\right]+   \frac{A_2}{\sqrt{2\pi}\Delta_2} \exp\left[-\frac{1}{2}\left(\frac{\ln k -\ln k_{p_2}}{\Delta_2}\right)^2\right].
	\end{equation}
	
	We conducted a Bayesian analysis of the NANOGrav 15 yrs data to investigate the parameterization of the power spectrum of the primordial curvature perturbation, as described by Eqs.~(\ref{eq:delta}-\ref{eq:dln}). 
	In our analysis, we utilized the 14 frequency bins reported in \cite{NANOGrav:2023gor, NANOGrav:2023hvm}  to fit the posterior distributions of the model parameters. The {\tt{Bilby}} code \cite{Ashton:2018jfp}  was employed for the analysis, utilizing the {\tt{dynesty}} algorithm for nested sampling  \cite{NestedSampling}.
	The log-likelihood function was constructed by evaluating the energy density of  SIGWs at the 14 specific frequency bins. Subsequently, we computed the sum of the logarithm of the probability density functions obtained from 14 independent kernel density estimates corresponding to these frequency values \cite{Moore:2021ibq, Lamb:2023jls, EPTA:2023hof}. The equation for the likelihood function is presented as
	\begin{equation}
		\mathcal{L}(\mathbf{\Theta})=\prod_{i=1}^{14} \mathcal{L}_i\left(\Omega_{\mathrm{GW}}\left(f_i, \mathbf{\Theta} \right)\right),
	\end{equation}
	where $\boldsymbol{\Theta}$ is the collection of parameters for $\delta$-function, box, lognormal, broken power law, and double lognormal models. These parameters and their priors are shown in Table~\ref{tab:prior_posterior}.
	
\begin{figure}[htp]
		\centering
		\includegraphics[width=0.7\textwidth]{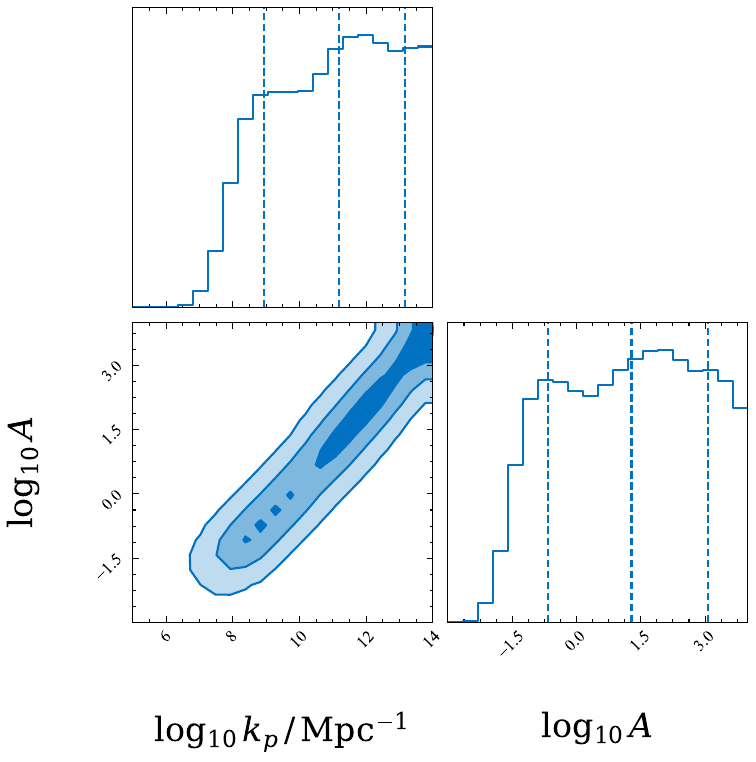}
		\caption{The posteriors of the parameters for the $\delta$-function parameterization \eqref{eq:delta}.}\label{fig:delta}
	\end{figure}

\begin{figure}[htp]
		\centering
		\includegraphics[width=0.8\textwidth]{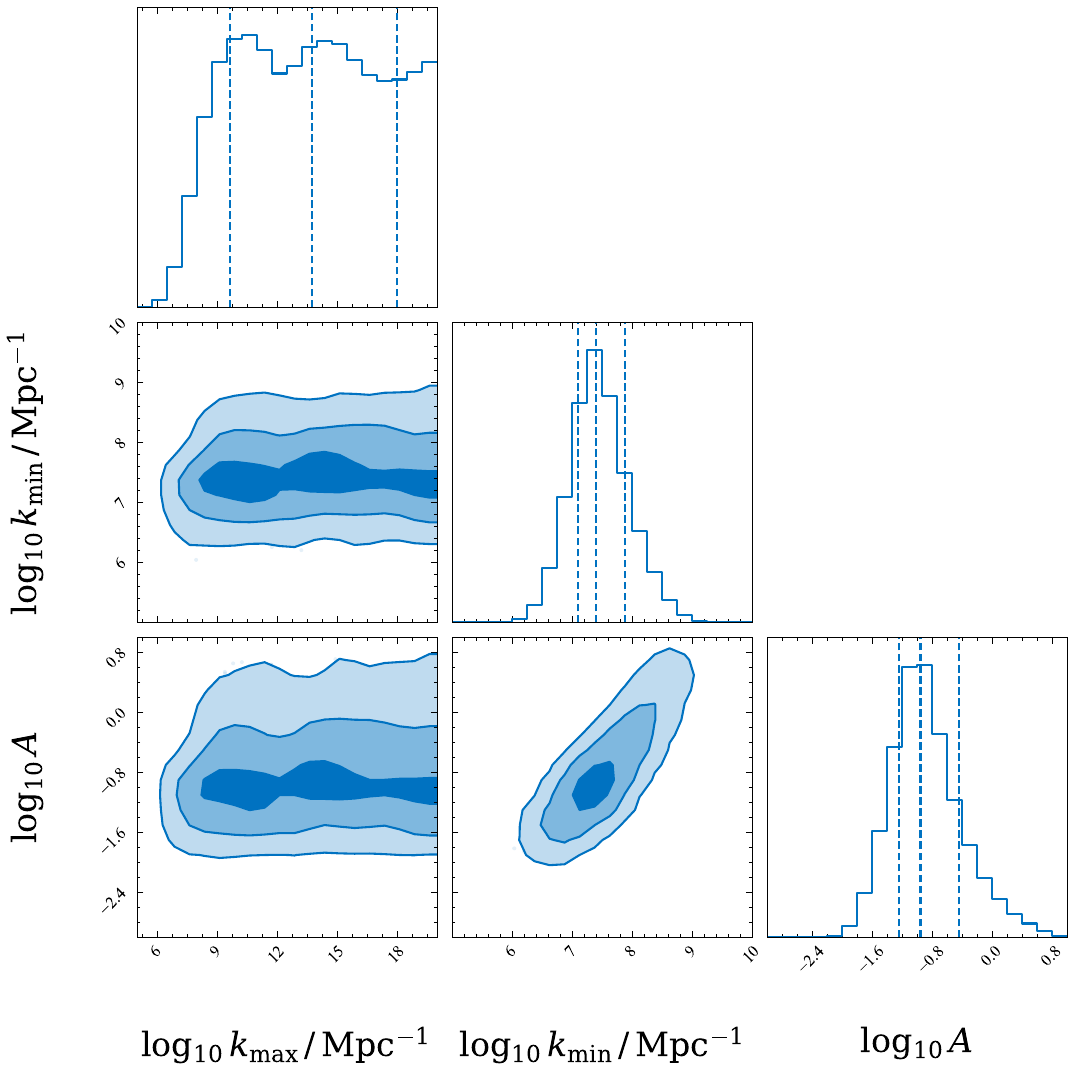}
		\caption{The posteriors of the parameters  for the box parameterization \eqref{eq:box}.} \label{fig:box}
	\end{figure}

\begin{figure}[htp]
		\centering
		\includegraphics[width=0.8\textwidth]{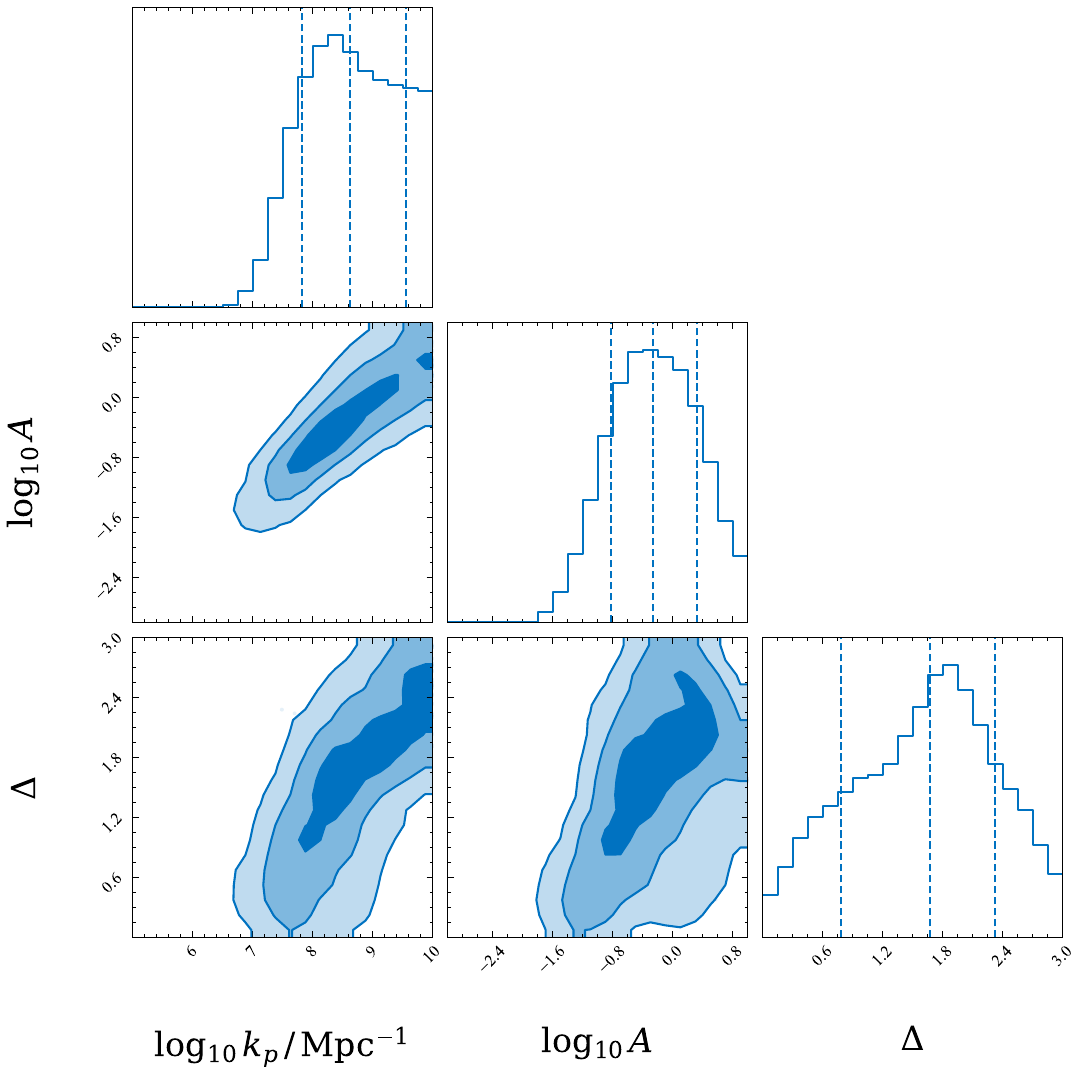}
		\caption{The posteriors on the parameters  of the lognormal  parameterization \eqref{eq:ln}.} \label{fig:ln}
	\end{figure}

\begin{figure}[htp]
		\centering
		\includegraphics[width=0.9\textwidth]{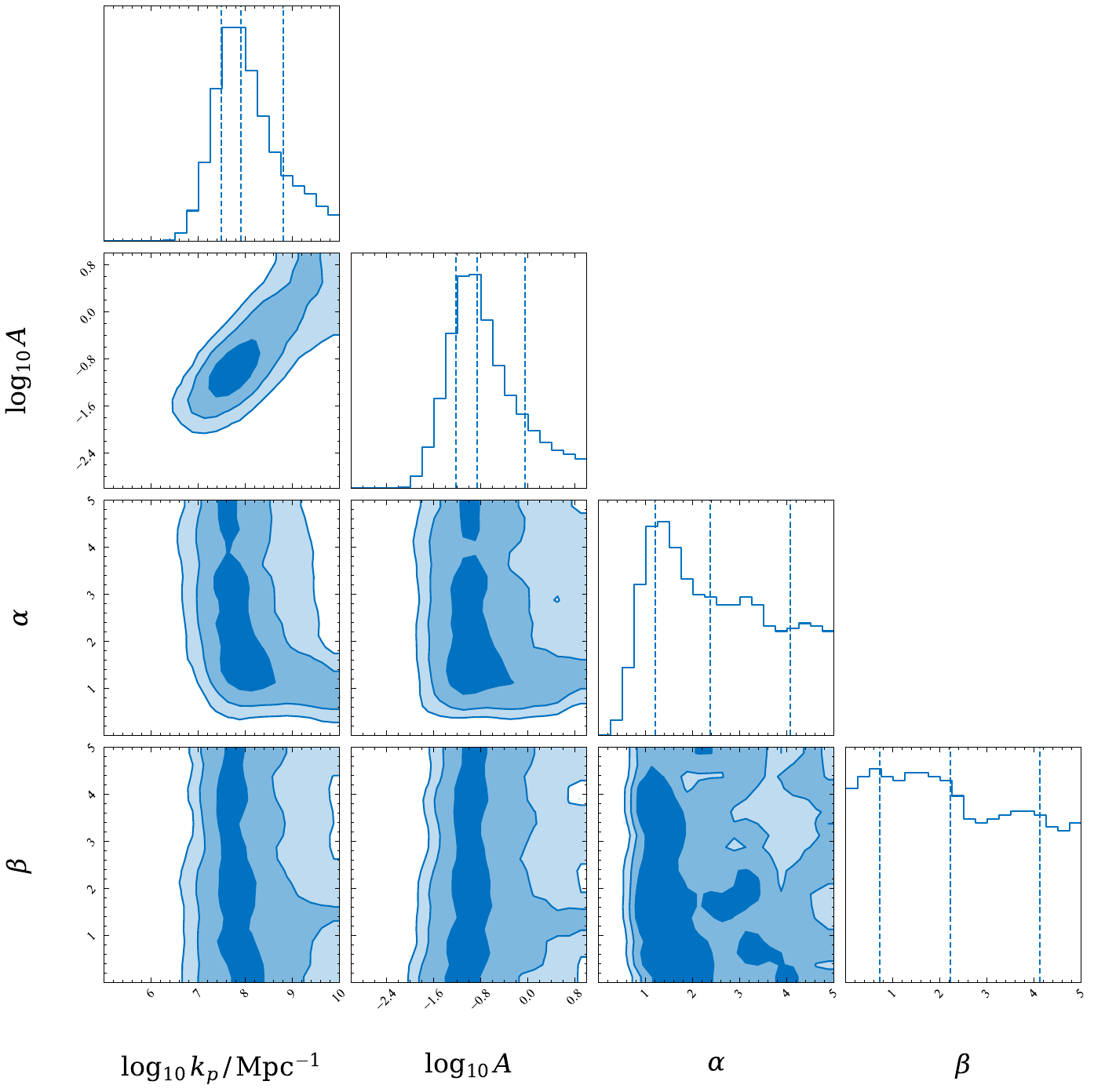}
		\caption{The posteriors on the parameters of the broken power law parameterization \eqref{eq:bp}.} \label{fig:bp}
	\end{figure}

\begin{figure}[htp]
		\centering
		\includegraphics[width=\textwidth]{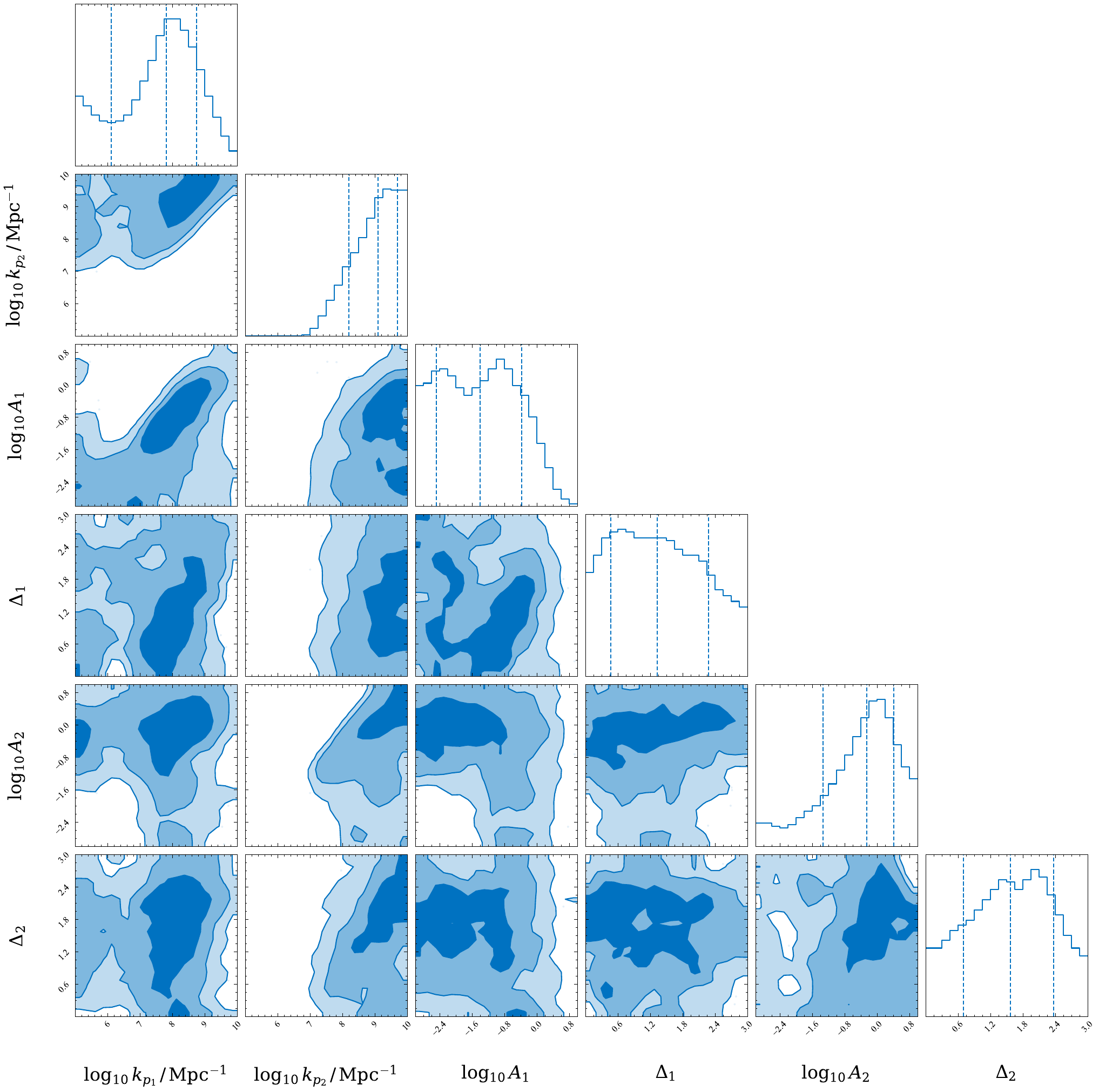}
		\caption{ The posteriors on the parameters  of the double lognormal  parameterization \eqref{eq:dln}.}\label{fig:dln}
	\end{figure}

	\begin{table}[htp]
		\centering
		\begin{tabular}{lclcr}
			\hline \hline  \\
			Model & Parameters & Prior & Posterior & ${\rm{ln}}\mathcal{B}$ \\
			\hline 
			& $ {\rm {log_{10}}}\,A$ &  $U[-3, 4]$ &  $1.28^{+1.79}_{-1.95}$ \\
			\textbf{$\delta$-function} & ${\rm log_{10}}\,k_p\,/\,{\rm Mpc^{-1} }$ &  $U[5,14]$  &  $11.19^{+1.99}_{-2.26}$   &  $0.29$  \\
			\hline
			& $ {\rm log_{10}}\,A$ & $U[-3, 1]$ & $-0.95^{+0.52}_{-0.28}$ \\
			\textbf{Box} & ${\rm log_{10}}\,k_{\mathrm{min}}\,/\,{\rm Mpc^{-1} }$ & $U[5,10]$ & $7.40^{+0.47}_{-0.31}$ & $0.26$\\
			& ${\rm log_{10}}\,k_{\mathrm{max}}\,/\, {\rm Mpc^{-1} }$ & $U[5,20]$ & $13.77^{+4.24}_{-4.11}$ \\
			\hline 
			& ${\rm log_{10}}\, A$ & $U[-3, 1]$ & $-0.88^{+0.81}_{-0.36}$\\
			& $\alpha$ & $U[0, 5]$ & $2.33^{+1.72}_{-1.14}$ \\
			\textbf{Broken power law} & $\beta$ & $U[0,5]$ & $2.22^{+1.87}_{-1.53}$ & $0.46$ \\
			& ${\rm log_{10}} \, k_p\,/\, {\rm Mpc^{-1} }$ & $U[5,10]$ & $7.91^{+0.88}_{-0.44}$ \\
			\hline 
			& $ {\rm {log_{10}}}\,A$ & $U[-3, 1]$ & $-0.26^{+0.60}_{-0.58}$ \\
			\textbf{Lognormal} & $\Delta$ & $U[0, 3]$ & $1.66^{+0.65}_{-0.87}$ & $0.45$ \\
			& ${\rm log_{10}} \, k_p\,/\, {\rm Mpc^{-1} }$ & $U[5,10]$ & $8.61^{+0.93}_{-0.79}$ \\
			\hline 
			& $ {\rm {log_{10}}}\,A_1$ & $U[-3, 1]$ & $-0.34^{+0.64}_{-0.61}$ \\
			& $\Delta_1$ & $U[0, 3]$ & $1.32^{+0.77}_{-0.77}$ \\
			\textbf{Double lognormal}& ${\rm log_{10}} \, k_{p_1}\,/\, {\rm Mpc^{-1} }$ & $U[5,10]$ & $8.21^{+0.92}_{-0.66}$ & $0$\\
			& $ {\rm {log_{10}}}\,A_2$ & $U[-3, 1]$ & $ -0.84^{+1.25}_{-1.49}$  \\
			& $\Delta_2$ & $U[0, 3]$ & $1.67^{+0.94}_{-1.09}$ \\
			&  ${\rm log_{10}} \, k_{p_2}\,/\, {\rm Mpc^{-1} }$ & $U[5,20]$ & $14.40^{+3.75}_{-3.90}$ \\
			\hline 
			\hline 
		\end{tabular}
		\caption{The priors, maximum posterior values, 1-$\sigma$ credible intervals bounds of posteriors and Bayes factor for $\delta$-function, box, lognormal, broken power law, and double lognormal model of the primordial curvature power spectrum using NANOGrav 15-yr data set. Here, we set the double lognormal model as the fiducial model.}\label{tab:prior_posterior}
	\end{table}
	We divide these models into two categories. The first one is single-peak power spectrum models, including $\delta$-function  \eqref{eq:delta}, box \eqref{eq:box}, lognormal \eqref{eq:ln} and broken power law model \eqref{eq:bp}, while the second one is double-peak model, including double lognormal model \eqref{eq:dln}. The posterior distributions for the parameters in Eqs.~(\ref{eq:delta}-\ref{eq:dln})  are depicted in Figures~\ref{fig:delta}-\ref{fig:dln}, respectively. We summarize the mean values and 1-$\sigma$ confidence intervals for parameters of these models in Table~\ref{tab:prior_posterior}.
	
	When comparing the results of the double-peak lognormal primordial curvature power spectrum with the single-peak models using $\delta$, box, lognormal, and broken power law forms, the Bayesian analysis yields no support in favor of the single-peak models with respective Bayes factors of  $\ln \mathcal{B}= 0.29$, $\ln \mathcal{B}=0.26$, $\ln \mathcal{B} =0.46$, and $\ln \mathcal{B} =0.45$. Thus, the PTAs data show no significant evidence for or against the single-peak primordial curvature power spectrum over the double-peak primordial curvature power spectrum.  
	Due to the very close values of logarithmic evidence, it is also difficult to favor which single-peak model provides a better fit.

\begin{figure}[htp]
		\centering
		\includegraphics[width=\textwidth]{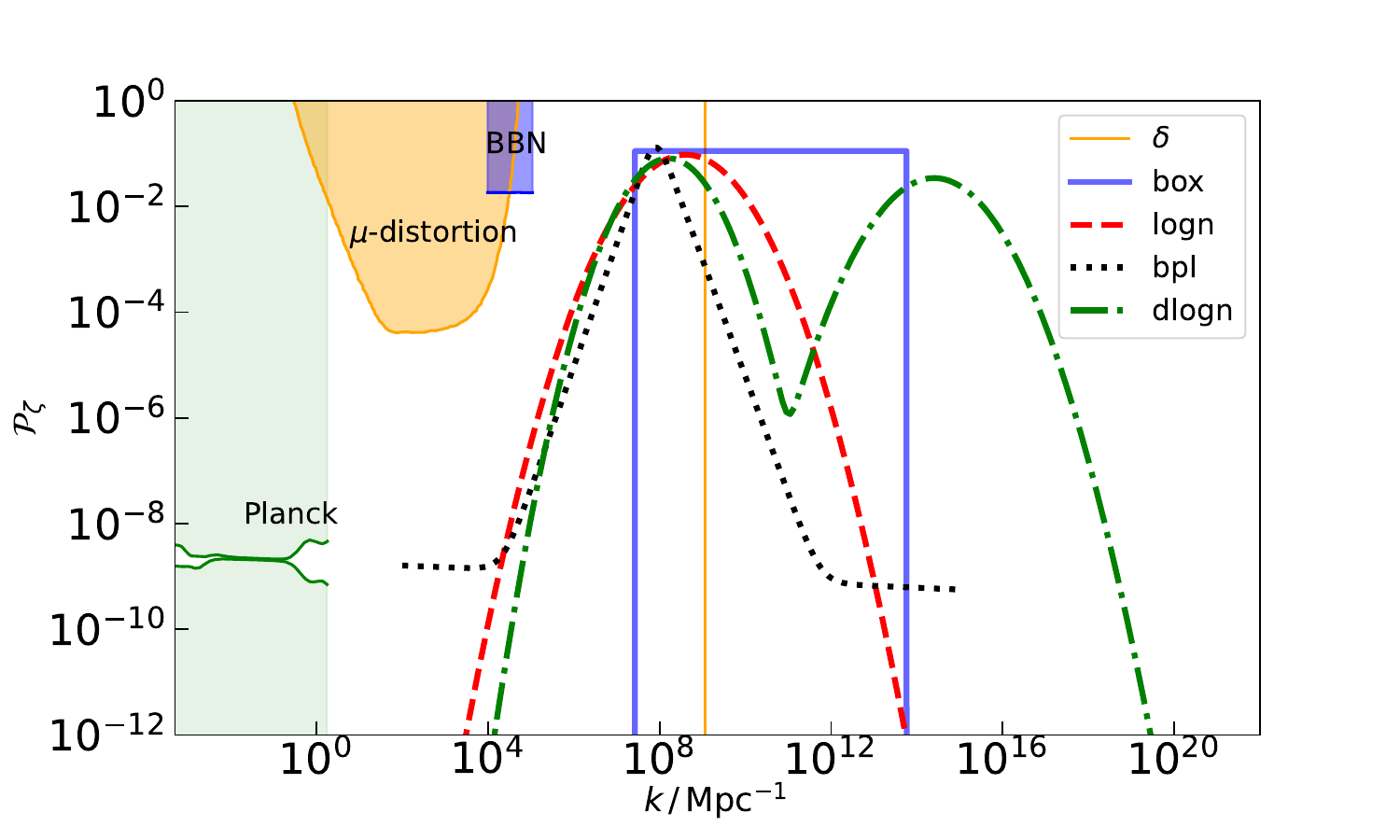}
		\caption{The power spectrum of the primordial curvature perturbation with  $\delta$ function, box, lognormal,  broken power law, and double lognormal parameterizations by choosing the best-fit parameters value in Table~\ref{tab:prior_posterior}. The labels ``$\delta$", ``box", ``logn", ``bpl", and ``dlogn" denote  $\delta$ function, box, lognormal,  broken power law, and double lognormal
			parameterizations, respectively.  		
			The light green shaded region is excluded by the CMB observations \cite{Planck:2018jri}.
			The  blue and orange regions are the
			constraints from  the effect on the ratio between neutron and proton
			during the Big Bang nucleosynthesis (BBN) \cite{Jeong:2014gna, Inomata:2016uip}
			and $\mu$-distortion of CMB \cite{Fixsen:1996nj,Chluba:2012we}, respectively.}\label{fig:powerspectrum}
	\end{figure}

\begin{figure}[htp]
		\centering
		\includegraphics[width=\textwidth]{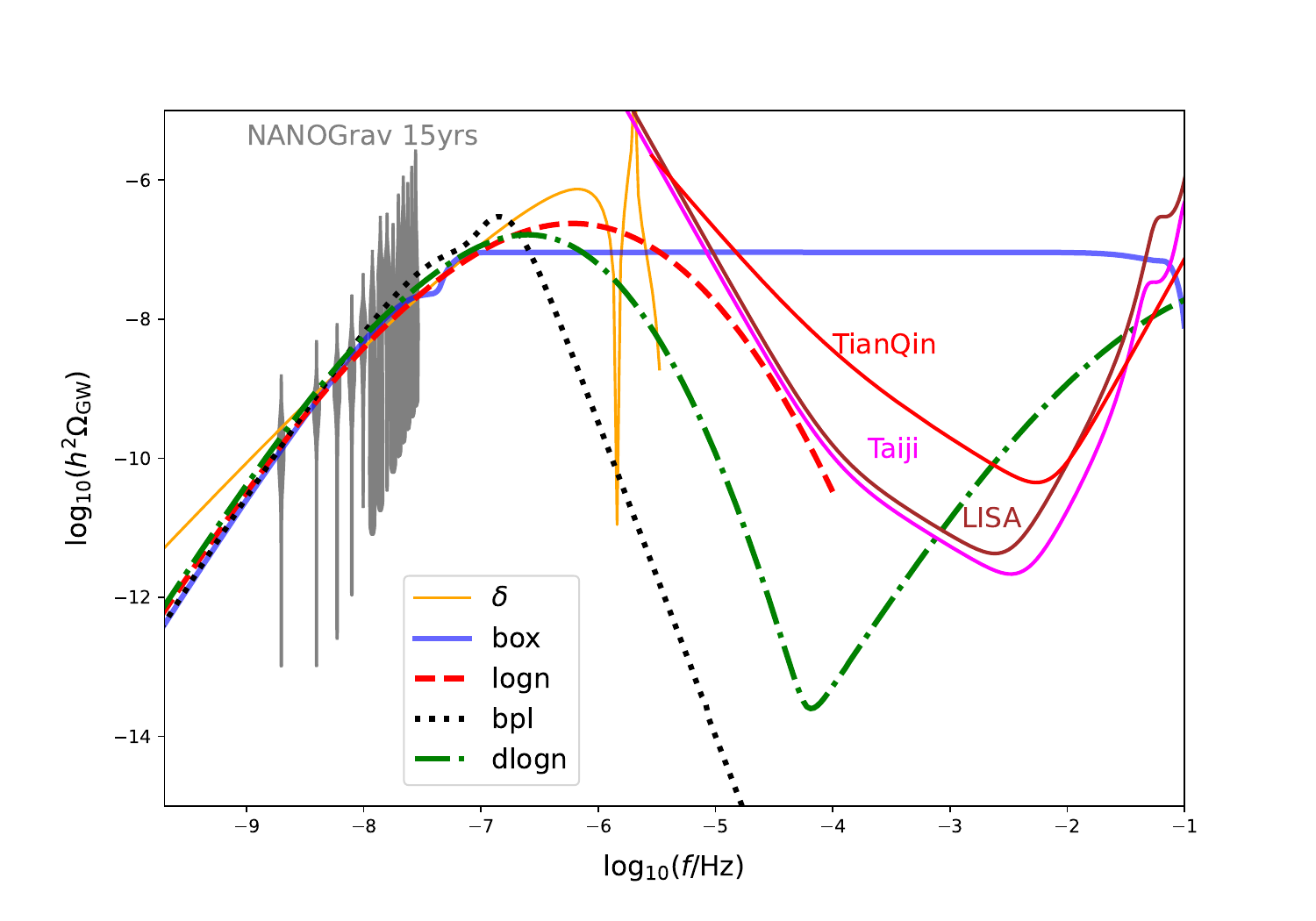}
		\caption{The energy density of SIGWs from the primordial curvature power spectrum parameterizations as displayed in Figure \ref{fig:powerspectrum}.   The labels ``$\delta$", ``box", ``logn", ``bpl", and ``dlogn" denote  $\delta$ function, box, lognormal,  broken power law, and double lognormal
			parameterizations of the primordial curvature power spectrum, respectively.   The grey regions represent the results form NANOGrav 15-yr data set~\cite{NANOGrav:2023gor, NANOGrav:2023hvm}, the red curve denotes the TianQin sensitivity curve \cite{Luo:2015ght}, the magenta   curve shows the Taiji sensitivity curve \cite{Hu:2017mde}, the brown  curve shows the LISA sensitivity curve \cite{Audley:2017drz}.}\label{fig:sigw}
	\end{figure}

	After obtaining the best-fit values from posteriors, we present the power spectrum of the primordial curvature perturbations in Figure \ref{fig:powerspectrum} and the corresponding  SIGWs in Figure \ref{fig:sigw}. In Figure \ref{fig:powerspectrum}, the orange thin solid line, blue thick solid line, red dashed line, black dotted line, and green dash-dotted line denote the primordial curvature power spectrum with the $\delta$-function, box, lognormal, broken power law, and double-lognormal parameterizations,  respectively.  The peak scale of these parameterizations is around $k_p\sim 10^{8} ~{\rm Mpc^{-1}}$, and the amplitude of the primordial curvature power spectrum of these parameterizations at the peak is around $A\sim 0.1$. 
	
	In Figure \ref{fig:sigw},  the orange thin solid line, blue thick solid line, red dashed line, black dotted line, and green dash-dotted line represent the energy density of the SIGW  from the primordial curvature power spectrum with the $\delta$-function, box, lognormal, broken power law, and double-lognormal parameterizations, respectively. If the PTAs data indeed arises from the SIGWs,  this PTAs signal can also be detected by space-based detectors in the future. And the parameterizations of the primordial curvature power spectrum can also be distinguished by the space-based detectors.

	\section{\label{discussion}Discussion and conclusion}
	The stochastic signal detected by the NANOGrav, PPTA,  EPTA, and CPTA collaborations points to the GW origin and can be explained by the SIGWs, where the scalar perturbations are seeded from the primordial curvature perturbations. 
	To determine the SIGWs model that best fits the observed stochastic signal,  we explore both single-peak and double-peak parameterizations for the power spectrum of the primordial curvature perturbations.
	For the single-peak scenarios, we consider parameterizations using the $\delta$-function form, box form, lognormal form, and broken power law form. Additionally, in the double-peak scenario, we employ the double lognormal form. 
	The best-fit values for the scale and amplitude of the primordial curvature perturbations at the peak, obtained from these five parameterizations, are approximately $k_p \sim 10^{8} ~ {\rm Mpc^{-1}}$ and $A\sim 0.1$. More specifically, they are  $\log_{10}A = 1.28^{+1.79}_{-1.95}$  and  $\log_{10} k_p/ \mathrm{Mpc}^{-1} =11.19^{+1.99}_{-2.26}$  for the $\delta$ function model;  $\log_{10}A = -0.95^{+0.52}_{-0.28}$, $\log_{10} k_{\mathrm{min}}/ \mathrm{Mpc}^{-1} =7.40^{+0.47}_{-0.31}$ and $\log_{10} k_{\mathrm{max}}/ \mathrm{Mpc}^{-1} =13.77^{+4.24}_{-4.11}$ for the box model; $\log_{10}A = -0.88^{+0.81}_{-0.36}$, $\alpha = 2.33^{+1.72}_{-1.14}$, $\beta = 2.22^{+1.87}_{-1.53}$ and  $\log_{10} k_p/ \mathrm{Mpc}^{-1} =7.91^{+0.88}_{-0.44}$ for the Broken power law model; $\log_{10}A = -0.26^{+0.60}_{-0.58}$, $\Delta=1.66^{+0.65}_{-0.87}$, and $\log_{10} k_p/ \mathrm{Mpc}^{-1} =8.61^{+0.93}_{-0.79}$ for the lognormal model; $\log_{10}A_1 = -0.34^{+0.64}_{-0.61}$, $\Delta_1=1.32^{+0.77}_{-0.77}$,   $\log_{10} k_{p_1}/ \mathrm{Mpc}^{-1} =8.21^{+0.92}_{-0.66}$, $\log_{10}A_2 = -0.84^{+1.25}_{-1.49}$, $\Delta_2=1.67^{+0.94}_{-1.09}$,  $\log_{10} k_{p_2}/ \mathrm{Mpc}^{-1} =14.40^{+3.75}_{-3.90}$ for the double lognormal model. 
	Comparing the results with the double-peak scenarios, the Bayesian analysis provides no support in favor of the single-peak models, with respective  Bayes factors of  $\ln \mathcal{B}= 0.29$, $\ln \mathcal{B}=0.26$, $\ln \mathcal{B} =0.46$, and $\ln \mathcal{B} =0.45$  for the $\delta$-function, box, lognormal, and broken power law forms, respectively. 
	If the stochastic signal observed by the  PTAs indeed originates from SIGWs, it may also be detectable by space-based gravitational wave detectors in the future, potentially allowing for the distinction between different types of SIGWs. Although our analysis in this paper focuses on the double-peak model, our conclusion can be extended to multi-peak models.

	When considering the PBHs generated accompanied by SIGWs, there may be a challenge of potentially producing an excessive number of PBHs. One potential solution to address this concern is to take into account non-Gaussianities~\cite{Franciolini:2023pbf,Liu:2023ymk}. This is a critical matter in the context of SIGWs attempting to explain the NANOGrav 15-year data set, and we leave it to future research. 
	
	In conclusion, the recent gravitational wave background signal can be explained by SIGWs, without preference for a single peak in the primordial curvature power spectrum over a multi-peak configuration.


	
	\acknowledgments
	We thank Xiao-Jing Liu for useful discussions. 
	ZQY is supported by the National Natural Science Foundation of China under Grant No.~12305059.
	ZY is supported by the National Natural Science
	Foundation of China under Grant No. 12205015 and the supporting fund for young researcher
	of Beijing Normal University under Grant No. 28719/310432102.


\begin{thebibliography}{100}
		
		\bibitem{NANOGrav:2023hde}
		{\scshape NANOGrav} collaboration, \emph{{The NANOGrav 15 yr Data Set:
				Observations and Timing of 68 Millisecond Pulsars}},
		\href{https://doi.org/10.3847/2041-8213/acda9a}{\emph{Astrophys. J. Lett.}
			{\bfseries 951} (2023) L9}
		[\href{https://arxiv.org/abs/2306.16217}{{\ttfamily 2306.16217}}].
		
		\bibitem{NANOGrav:2023gor}
		{\scshape NANOGrav} collaboration, \emph{{The NANOGrav 15 yr Data Set: Evidence
				for a Gravitational-wave Background}},
		\href{https://doi.org/10.3847/2041-8213/acdac6}{\emph{Astrophys. J. Lett.}
			{\bfseries 951} (2023) L8}
		[\href{https://arxiv.org/abs/2306.16213}{{\ttfamily 2306.16213}}].
		
		\bibitem{Zic:2023gta}
		A.~Zic et~al., \emph{{The Parkes Pulsar Timing Array Third Data Release}},
		\href{https://arxiv.org/abs/2306.16230}{{\ttfamily 2306.16230}}.
		
		\bibitem{Reardon:2023gzh}
		D.J.~Reardon et~al., \emph{{Search for an Isotropic Gravitational-wave
				Background with the Parkes Pulsar Timing Array}},
		\href{https://doi.org/10.3847/2041-8213/acdd02}{\emph{Astrophys. J. Lett.}
			{\bfseries 951} (2023) L6}
		[\href{https://arxiv.org/abs/2306.16215}{{\ttfamily 2306.16215}}].
		
		\bibitem{Antoniadis:2023lym}
		{\scshape EPTA} collaboration, \emph{{The second data release from the European
				Pulsar Timing Array I. The dataset and timing analysis}},
		\href{https://doi.org/10.1051/0004-6361/202346841}{\emph{Astron. Astrophys.}
			{\bfseries 678} (2023) A48}
		[\href{https://arxiv.org/abs/2306.16224}{{\ttfamily 2306.16224}}].
		
		\bibitem{Antoniadis:2023ott}
		{\scshape EPTA} collaboration, \emph{{The second data release from the European
				Pulsar Timing Array III. Search for gravitational wave signals}},
		\href{https://doi.org/10.1051/0004-6361/202346844}{\emph{Astron. Astrophys.}
			{\bfseries 678} (2023) A50}
		[\href{https://arxiv.org/abs/2306.16214}{{\ttfamily 2306.16214}}].
		
		\bibitem{Xu:2023wog}
		H.~Xu et~al., \emph{{Searching for the Nano-Hertz Stochastic Gravitational Wave
				Background with the Chinese Pulsar Timing Array Data Release I}},
		\href{https://doi.org/10.1088/1674-4527/acdfa5}{\emph{Res. Astron.
				Astrophys.} {\bfseries 23} (2023) 075024}
		[\href{https://arxiv.org/abs/2306.16216}{{\ttfamily 2306.16216}}].
		
		\bibitem{NANOGrav:2023hvm}
		{\scshape NANOGrav} collaboration, \emph{{The NANOGrav 15 yr Data Set: Search
				for Signals from New Physics}},
		\href{https://doi.org/10.3847/2041-8213/acdc91}{\emph{Astrophys. J. Lett.}
			{\bfseries 951} (2023) L11}
		[\href{https://arxiv.org/abs/2306.16219}{{\ttfamily 2306.16219}}].
		
		\bibitem{Antoniadis:2023xlr}
		{\scshape EPTA} collaboration, \emph{{The second data release from the European
				Pulsar Timing Array: V. Implications for massive black holes, dark matter and
				the early Universe}},  \href{https://arxiv.org/abs/2306.16227}{{\ttfamily
				2306.16227}}.
		
		\bibitem{Franciolini:2023pbf}
		G.~Franciolini, A.~Iovino, Junior., V.~Vaskonen and H.~Veermae, \emph{{The
				recent gravitational wave observation by pulsar timing arrays and primordial
				black holes: the importance of non-gaussianities}},
		\href{https://arxiv.org/abs/2306.17149}{{\ttfamily 2306.17149}}.
		
		\bibitem{Liu:2023ymk}
		L.~Liu, Z.-C.~Chen and Q.-G.~Huang, \emph{{Implications for the non-Gaussianity
				of curvature perturbation from pulsar timing arrays}},
		\href{https://arxiv.org/abs/2307.01102}{{\ttfamily 2307.01102}}.
		
		\bibitem{Vagnozzi:2023lwo}
		S.~Vagnozzi, \emph{{Inflationary interpretation of the stochastic gravitational
				wave background signal detected by pulsar timing array experiments}},
		\href{https://doi.org/10.1016/j.jheap.2023.07.001}{\emph{JHEAp} {\bfseries
				39} (2023) 81} [\href{https://arxiv.org/abs/2306.16912}{{\ttfamily
				2306.16912}}].
		
		\bibitem{Cai:2023dls}
		Y.-F.~Cai, X.-C.~He, X.~Ma, S.-F.~Yan and G.-W.~Yuan, \emph{{Limits on
				scalar-induced gravitational waves from the stochastic background by pulsar
				timing array observations}},
		\href{https://arxiv.org/abs/2306.17822}{{\ttfamily 2306.17822}}.
		
		\bibitem{Wang:2023ost}
		S.~Wang, Z.-C.~Zhao, J.-P.~Li and Q.-H.~Zhu, \emph{{Implications of Pulsar
				Timing Array Data for Scalar-Induced Gravitational Waves and Primordial Black
				Holes: Primordial Non-Gaussianity $f_{\mathrm{NL}}$ Considered}},
		\href{https://arxiv.org/abs/2307.00572}{{\ttfamily 2307.00572}}.
		
		\bibitem{Yi:2023mbm}
		Z.~Yi, Q.~Gao, Y.~Gong, Y.~Wang and F.~Zhang, \emph{{The waveform of the scalar
				induced gravitational waves in light of Pulsar Timing Array data}},
		\href{https://arxiv.org/abs/2307.02467}{{\ttfamily 2307.02467}}.
		
		\bibitem{Bi:2023tib}
		Y.-C.~Bi, Y.-M.~Wu, Z.-C.~Chen and Q.-G.~Huang, \emph{{Implications for the
				Supermassive Black Hole Binaries from the NANOGrav 15-year Data Set}},
		\href{https://arxiv.org/abs/2307.00722}{{\ttfamily 2307.00722}}.
		
		\bibitem{Wu:2023hsa}
		Y.-M.~Wu, Z.-C.~Chen and Q.-G.~Huang, \emph{{Cosmological Interpretation for
				the Stochastic Signal in Pulsar Timing Arrays}},
		\href{https://arxiv.org/abs/2307.03141}{{\ttfamily 2307.03141}}.
		
		\bibitem{Zhu:2023faa}
		S.~Wang, Z.-C.~Zhao and Q.-H.~Zhu, \emph{{Constraints On Scalar-Induced
				Gravitational Waves Up To Third Order From Joint Analysis of BBN, CMB, And
				PTA Data}},  \href{https://arxiv.org/abs/2307.03095}{{\ttfamily 2307.03095}}.
		
		\bibitem{Franciolini:2023wjm}
		G.~Franciolini, D.~Racco and F.~Rompineve, \emph{{Footprints of the QCD
				Crossover on Cosmological Gravitational Waves at Pulsar Timing Arrays}},
		\href{https://arxiv.org/abs/2306.17136}{{\ttfamily 2306.17136}}.
		
		\bibitem{Jin:2023wri}
		J.-H.~Jin, Z.-C.~Chen, Z.~Yi, Z.-Q.~You, L.~Liu and Y.~Wu, \emph{{Confronting
				sound speed resonance with pulsar timing arrays}},
		\href{https://doi.org/10.1088/1475-7516/2023/09/016}{\emph{JCAP} {\bfseries
				09} (2023) 016} [\href{https://arxiv.org/abs/2307.08687}{{\ttfamily
				2307.08687}}].
		
		\bibitem{Liu:2023pau}
		L.~Liu, Z.-C.~Chen and Q.-G.~Huang, \emph{{Probing the equation of state of the
				early Universe with pulsar timing arrays}},
		\href{https://arxiv.org/abs/2307.14911}{{\ttfamily 2307.14911}}.
		
		\bibitem{Yi:2023npi}
		Z.~Yi, Z.-Q.~You, Y.~Wu, Z.-C.~Chen and L.~Liu, \emph{{Exploring the NANOGrav
				Signal and Planet-mass Primordial Black Holes through Higgs Inflation}},
		\href{https://arxiv.org/abs/2308.14688}{{\ttfamily 2308.14688}}.
		
		\bibitem{Yi:2023tdk}
		Z.~Yi, Z.-Q.~You and Y.~Wu, \emph{{Model-independent reconstruction of the
				primordial curvature power spectrum from PTA data}},
		\href{https://arxiv.org/abs/2308.05632}{{\ttfamily 2308.05632}}.
		
		\bibitem{You:2023rmn}
		Z.-Q.~You, Z.~Yi and Y.~Wu, \emph{{Constraints on primordial curvature power
				spectrum with pulsar timing arrays}},
		\href{https://arxiv.org/abs/2307.04419}{{\ttfamily 2307.04419}}.
		
		\bibitem{Wu:2023rib}
		Y.-M.~Wu, Z.-C.~Chen, Y.-C.~Bi and Q.-G.~Huang, \emph{{Constraining the
				Graviton Mass with the NANOGrav 15-Year Data Set}},
		\href{https://arxiv.org/abs/2310.07469}{{\ttfamily 2310.07469}}.
		
		\bibitem{Bi:2023ewq}
		Y.-C.~Bi, Y.-M.~Wu, Z.-C.~Chen and Q.-G.~Huang, \emph{{Constraints on the
				velocity of gravitational waves from NANOGrav 15-year data set}},
		\href{https://arxiv.org/abs/2310.08366}{{\ttfamily 2310.08366}}.
		
		\bibitem{Chen:2023uiz}
		Z.-C.~Chen, Y.-M.~Wu, Y.-C.~Bi and Q.-G.~Huang, \emph{{Search for Non-Tensorial
				Gravitational-Wave Backgrounds in the NANOGrav 15-Year Data Set}},
		\href{https://arxiv.org/abs/2310.11238}{{\ttfamily 2310.11238}}.
		
		\bibitem{Chen:2023swcs}
		L.~Liu, Y.~Wu and Z.-C.~Chen, \emph{{Simultaneously probing the sound speed and
				equation of state of the early Universe with pulsar timing arrays}},
		\href{https://arxiv.org/abs/2310.16500}{{\ttfamily 2310.16500}}.
		
		\bibitem{Zeldovich:1967lct}
		Y.B.~Zel'dovich and I.D.~Novikov, \emph{{The Hypothesis of Cores Retarded
				during Expansion and the Hot Cosmological Model}}, {\emph{Soviet Astron. AJ
				(Engl. Transl. ),} {\bfseries 10} (1967) 602}.
		
		\bibitem{Hawking:1971ei}
		S.~Hawking, \emph{{Gravitationally collapsed objects of very low mass}},
		\href{https://doi.org/10.1093/mnras/152.1.75}{\emph{Mon. Not. Roy. Astron.
				Soc.} {\bfseries 152} (1971) 75}.
		
		\bibitem{Carr:1974nx}
		B.J.~Carr and S.W.~Hawking, \emph{{Black holes in the early Universe}},
		\href{https://doi.org/10.1093/mnras/168.2.399}{\emph{Mon. Not. Roy. Astron.
				Soc.} {\bfseries 168} (1974) 399}.
		
		\bibitem{Chen:2018czv}
		Z.-C.~Chen and Q.-G.~Huang, \emph{{Merger Rate Distribution of
				Primordial-Black-Hole Binaries}},
		\href{https://doi.org/10.3847/1538-4357/aad6e2}{\emph{Astrophys. J.}
			{\bfseries 864} (2018) 61}
		[\href{https://arxiv.org/abs/1801.10327}{{\ttfamily 1801.10327}}].
		
		\bibitem{Chen:2018rzo}
		Z.-C.~Chen, F.~Huang and Q.-G.~Huang, \emph{{Stochastic Gravitational-wave
				Background from Binary Black Holes and Binary Neutron Stars and Implications
				for LISA}}, \href{https://doi.org/10.3847/1538-4357/aaf581}{\emph{Astrophys.
				J.} {\bfseries 871} (2019) 97}
		[\href{https://arxiv.org/abs/1809.10360}{{\ttfamily 1809.10360}}].
		
		\bibitem{Liu:2018ess}
		L.~Liu, Z.-K.~Guo and R.-G.~Cai, \emph{{Effects of the surrounding primordial
				black holes on the merger rate of primordial black hole binaries}},
		\href{https://doi.org/10.1103/PhysRevD.99.063523}{\emph{Phys. Rev. D}
			{\bfseries 99} (2019) 063523}
		[\href{https://arxiv.org/abs/1812.05376}{{\ttfamily 1812.05376}}].
		
		\bibitem{Liu:2019rnx}
		L.~Liu, Z.-K.~Guo and R.-G.~Cai, \emph{{Effects of the merger history on the
				merger rate density of primordial black hole binaries}},
		\href{https://doi.org/10.1140/epjc/s10052-019-7227-0}{\emph{Eur. Phys. J. C}
			{\bfseries 79} (2019) 717}
		[\href{https://arxiv.org/abs/1901.07672}{{\ttfamily 1901.07672}}].
		
		\bibitem{Chen:2019irf}
		Z.-C.~Chen and Q.-G.~Huang, \emph{{Distinguishing Primordial Black Holes from
				Astrophysical Black Holes by Einstein Telescope and Cosmic Explorer}},
		\href{https://doi.org/10.1088/1475-7516/2020/08/039}{\emph{JCAP} {\bfseries
				08} (2020) 039} [\href{https://arxiv.org/abs/1904.02396}{{\ttfamily
				1904.02396}}].
		
		\bibitem{Liu:2020cds}
		L.~Liu, Z.-K.~Guo, R.-G.~Cai and S.P.~Kim, \emph{{Merger rate distribution of
				primordial black hole binaries with electric charges}},
		\href{https://doi.org/10.1103/PhysRevD.102.043508}{\emph{Phys. Rev. D}
			{\bfseries 102} (2020) 043508}
		[\href{https://arxiv.org/abs/2001.02984}{{\ttfamily 2001.02984}}].
		
		\bibitem{Liu:2020vsy}
		L.~Liu, O.~Christiansen, Z.-K.~Guo, R.-G.~Cai and S.P.~Kim,
		\emph{{Gravitational and electromagnetic radiation from binary black holes
				with electric and magnetic charges: Circular orbits on a cone}},
		\href{https://doi.org/10.1103/PhysRevD.102.103520}{\emph{Phys. Rev. D}
			{\bfseries 102} (2020) 103520}
		[\href{https://arxiv.org/abs/2008.02326}{{\ttfamily 2008.02326}}].
		
		\bibitem{Liu:2020bag}
		L.~Liu, O.~Christiansen, W.-H.~Ruan, Z.-K.~Guo, R.-G.~Cai and S.P.~Kim,
		\emph{{Gravitational and electromagnetic radiation from binary black holes
				with electric and magnetic charges: elliptical orbits on a cone}},
		\href{https://doi.org/10.1140/epjc/s10052-021-09849-4}{\emph{Eur. Phys. J. C}
			{\bfseries 81} (2021) 1048}
		[\href{https://arxiv.org/abs/2011.13586}{{\ttfamily 2011.13586}}].
		
		\bibitem{Wu:2020drm}
		Y.~Wu, \emph{{Merger history of primordial black-hole binaries}},
		\href{https://doi.org/10.1103/PhysRevD.101.083008}{\emph{Phys. Rev. D}
			{\bfseries 101} (2020) 083008}
		[\href{https://arxiv.org/abs/2001.03833}{{\ttfamily 2001.03833}}].
		
		\bibitem{Chen:2021nxo}
		Z.-C.~Chen, C.~Yuan and Q.-G.~Huang, \emph{{Confronting the primordial black
				hole scenario with the gravitational-wave events detected by LIGO-Virgo}},
		\href{https://doi.org/10.1016/j.physletb.2022.137040}{\emph{Phys. Lett. B}
			{\bfseries 829} (2022) 137040}
		[\href{https://arxiv.org/abs/2108.11740}{{\ttfamily 2108.11740}}].
		
		\bibitem{Liu:2022wtq}
		L.~Liu and S.P.~Kim, \emph{{Merger rate of charged black holes from the
				two-body dynamical capture}},
		\href{https://doi.org/10.1088/1475-7516/2022/03/059}{\emph{JCAP} {\bfseries
				03} (2022) 059} [\href{https://arxiv.org/abs/2201.02581}{{\ttfamily
				2201.02581}}].
		
		\bibitem{Chen:2022fda}
		Z.-C.~Chen, S.-S.~Du, Q.-G.~Huang and Z.-Q.~You, \emph{{Constraints on
				primordial-black-hole population and cosmic expansion history from GWTC-3}},
		\href{https://doi.org/10.1088/1475-7516/2023/03/024}{\emph{JCAP} {\bfseries
				03} (2023) 024} [\href{https://arxiv.org/abs/2205.11278}{{\ttfamily
				2205.11278}}].
		
		\bibitem{Chen:2022qvg}
		Z.-C.~Chen, S.P.~Kim and L.~Liu, \emph{{Gravitational and electromagnetic
				radiation from binary black holes with electric and magnetic charges:
				hyperbolic orbits on a cone}},
		\href{https://doi.org/10.1088/1572-9494/acce98}{\emph{Commun. Theor. Phys.}
			{\bfseries 75} (2023) 065401}
		[\href{https://arxiv.org/abs/2210.15564}{{\ttfamily 2210.15564}}].
		
		\bibitem{Liu:2022iuf}
		L.~Liu, Z.-Q.~You, Y.~Wu and Z.-C.~Chen, \emph{{Constraining the merger history
				of primordial-black-hole binaries from GWTC-3}},
		\href{https://doi.org/10.1103/PhysRevD.107.063035}{\emph{Phys. Rev. D}
			{\bfseries 107} (2023) 063035}
		[\href{https://arxiv.org/abs/2210.16094}{{\ttfamily 2210.16094}}].
		
		\bibitem{Zheng:2022wqo}
		L.-M.~Zheng, Z.~Li, Z.-C.~Chen, H.~Zhou and Z.-H.~Zhu, \emph{{Towards a
				reliable reconstruction of the power spectrum of primordial curvature
				perturbation on small scales from GWTC-3}},
		\href{https://doi.org/10.1016/j.physletb.2023.137720}{\emph{Phys. Lett. B}
			{\bfseries 838} (2023) 137720}
		[\href{https://arxiv.org/abs/2212.05516}{{\ttfamily 2212.05516}}].
		
		\bibitem{Yi:2022ymw}
		Z.~Yi and Q.~Fei, \emph{{Constraints on primordial curvature spectrum from
				primordial black holes and scalar-induced gravitational waves}},
		\href{https://doi.org/10.1140/epjc/s10052-023-11233-3}{\emph{Eur. Phys. J. C}
			{\bfseries 83} (2023) 82} [\href{https://arxiv.org/abs/2210.03641}{{\ttfamily
				2210.03641}}].
		
		\bibitem{Zhu:2018lif}
		X.-J.~Zhu, W.~Cui and E.~Thrane, \emph{{The minimum and maximum
				gravitational-wave background from supermassive binary black holes}},
		\href{https://doi.org/10.1093/mnras/sty2849}{\emph{Mon. Not. Roy. Astron.
				Soc.} {\bfseries 482} (2019) 2588}
		[\href{https://arxiv.org/abs/1806.02346}{{\ttfamily 1806.02346}}].
		
		\bibitem{Li:2019vlb}
		J.~Li, Z.-C.~Chen and Q.-G.~Huang, \emph{{Measuring the tilt of primordial
				gravitational-wave power spectrum from observations}},
		\href{https://doi.org/10.1007/s11433-019-9605-5}{\emph{Sci. China Phys. Mech.
				Astron.} {\bfseries 62} (2019) 110421}
		[\href{https://arxiv.org/abs/1907.09794}{{\ttfamily 1907.09794}}].
		
		\bibitem{Chen:2021wdo}
		Z.-C.~Chen, C.~Yuan and Q.-G.~Huang, \emph{{Non-tensorial gravitational wave
				background in NANOGrav 12.5-year data set}},
		\href{https://doi.org/10.1007/s11433-021-1797-y}{\emph{Sci. China Phys. Mech.
				Astron.} {\bfseries 64} (2021) 120412}
		[\href{https://arxiv.org/abs/2101.06869}{{\ttfamily 2101.06869}}].
		
		\bibitem{Wu:2021kmd}
		Y.-M.~Wu, Z.-C.~Chen and Q.-G.~Huang, \emph{{Constraining the Polarization of
				Gravitational Waves with the Parkes Pulsar Timing Array Second Data
				Release}}, \href{https://doi.org/10.3847/1538-4357/ac35cc}{\emph{Astrophys.
				J.} {\bfseries 925} (2022) 37}
		[\href{https://arxiv.org/abs/2108.10518}{{\ttfamily 2108.10518}}].
		
		\bibitem{Chen:2021ncc}
		Z.-C.~Chen, Y.-M.~Wu and Q.-G.~Huang, \emph{{Searching for isotropic stochastic
				gravitational-wave background in the international pulsar timing array second
				data release}}, \href{https://doi.org/10.1088/1572-9494/ac7cdf}{\emph{Commun.
				Theor. Phys.} {\bfseries 74} (2022) 105402}
		[\href{https://arxiv.org/abs/2109.00296}{{\ttfamily 2109.00296}}].
		
		\bibitem{Chen:2022azo}
		Z.-C.~Chen, Y.-M.~Wu and Q.-G.~Huang, \emph{{Search for the Gravitational-wave
				Background from Cosmic Strings with the Parkes Pulsar Timing Array Second
				Data Release}},
		\href{https://doi.org/10.3847/1538-4357/ac86cb}{\emph{Astrophys. J.}
			{\bfseries 936} (2022) 20}
		[\href{https://arxiv.org/abs/2205.07194}{{\ttfamily 2205.07194}}].
		
		\bibitem{PPTA:2022eul}
		{\scshape PPTA} collaboration, \emph{{Constraining ultralight vector dark
				matter with the Parkes Pulsar Timing Array second data release}},
		\href{https://doi.org/10.1103/PhysRevD.106.L081101}{\emph{Phys. Rev. D}
			{\bfseries 106} (2022) L081101}
		[\href{https://arxiv.org/abs/2210.03880}{{\ttfamily 2210.03880}}].
		
		\bibitem{IPTA:2023ero}
		{\scshape IPTA} collaboration, \emph{{Searching for continuous Gravitational
				Waves in the second data release of the International Pulsar Timing Array}},
		\href{https://doi.org/10.1093/mnras/stad812}{\emph{Mon. Not. Roy. Astron.
				Soc.} {\bfseries 521} (2023) 5077}
		[\href{https://arxiv.org/abs/2303.10767}{{\ttfamily 2303.10767}}].
		
		\bibitem{Wu:2023pbt}
		Y.-M.~Wu, Z.-C.~Chen and Q.-G.~Huang, \emph{{Search for stochastic
				gravitational-wave background from massive gravity in the NANOGrav 12.5-year
				dataset}}, \href{https://doi.org/10.1103/PhysRevD.107.042003}{\emph{Phys.
				Rev. D} {\bfseries 107} (2023) 042003}
		[\href{https://arxiv.org/abs/2302.00229}{{\ttfamily 2302.00229}}].
		
		\bibitem{Wu:2023dnp}
		Y.-M.~Wu, Z.-C.~Chen and Q.-G.~Huang, \emph{{Pulsar timing residual induced by
				ultralight tensor dark matter}},
		\href{https://doi.org/10.1088/1475-7516/2023/09/021}{\emph{JCAP} {\bfseries
				09} (2023) 021} [\href{https://arxiv.org/abs/2305.08091}{{\ttfamily
				2305.08091}}].
		
		\bibitem{InternationalPulsarTimingArray:2023mzf}
		{\scshape International Pulsar Timing Array} collaboration, \emph{{Comparing
				recent PTA results on the nanohertz stochastic gravitational wave
				background}},  \href{https://arxiv.org/abs/2309.00693}{{\ttfamily
				2309.00693}}.
		
		\bibitem{Chen:2023zkb}
		Z.-C.~Chen, Q.-G.~Huang, C.~Liu, L.~Liu, X.-J.~Liu, Y.~Wu et~al.,
		\emph{{Prospects for Taiji to detect a gravitational-wave background from
				cosmic strings}},  \href{https://arxiv.org/abs/2310.00411}{{\ttfamily
				2310.00411}}.
		
		\bibitem{tomita1967non}
		K.~Tomita, \emph{Non-linear theory of gravitational instability in the
			expanding universe}, {\emph{Progress of Theoretical Physics} {\bfseries 37}
			(1967) 831}.
		
		\bibitem{Saito:2008jc}
		R.~Saito and J.~Yokoyama, \emph{{Gravitational wave background as a probe of
				the primordial black hole abundance}},
		\href{https://doi.org/10.1103/PhysRevLett.102.161101}{\emph{Phys. Rev. Lett.}
			{\bfseries 102} (2009) 161101}
		[\href{https://arxiv.org/abs/0812.4339}{{\ttfamily 0812.4339}}].
		
		\bibitem{Young:2014ana}
		S.~Young, C.T.~Byrnes and M.~Sasaki, \emph{{Calculating the mass fraction of
				primordial black holes}},
		\href{https://doi.org/10.1088/1475-7516/2014/07/045}{\emph{JCAP} {\bfseries
				07} (2014) 045} [\href{https://arxiv.org/abs/1405.7023}{{\ttfamily
				1405.7023}}].
		
		\bibitem{Yuan:2019udt}
		C.~Yuan, Z.-C.~Chen and Q.-G.~Huang, \emph{{Probing
				primordial\textendash{}black-hole dark matter with scalar induced
				gravitational waves}},
		\href{https://doi.org/10.1103/PhysRevD.100.081301}{\emph{Phys. Rev. D}
			{\bfseries 100} (2019) 081301}
		[\href{https://arxiv.org/abs/1906.11549}{{\ttfamily 1906.11549}}].
		
		\bibitem{Yuan:2019wwo}
		C.~Yuan, Z.-C.~Chen and Q.-G.~Huang, \emph{{Log-dependent slope of scalar
				induced gravitational waves in the infrared regions}},
		\href{https://doi.org/10.1103/PhysRevD.101.043019}{\emph{Phys. Rev. D}
			{\bfseries 101} (2020) 043019}
		[\href{https://arxiv.org/abs/1910.09099}{{\ttfamily 1910.09099}}].
		
		\bibitem{Chen:2019xse}
		Z.-C.~Chen, C.~Yuan and Q.-G.~Huang, \emph{{Pulsar Timing Array Constraints on
				Primordial Black Holes with NANOGrav 11-Year Dataset}},
		\href{https://doi.org/10.1103/PhysRevLett.124.251101}{\emph{Phys. Rev. Lett.}
			{\bfseries 124} (2020) 251101}
		[\href{https://arxiv.org/abs/1910.12239}{{\ttfamily 1910.12239}}].
		
		\bibitem{Yuan:2019fwv}
		C.~Yuan, Z.-C.~Chen and Q.-G.~Huang, \emph{{Scalar induced gravitational waves
				in different gauges}},
		\href{https://doi.org/10.1103/PhysRevD.101.063018}{\emph{Phys. Rev. D}
			{\bfseries 101} (2020) 063018}
		[\href{https://arxiv.org/abs/1912.00885}{{\ttfamily 1912.00885}}].
		
		\bibitem{Ananda:2006af}
		K.N.~Ananda, C.~Clarkson and D.~Wands, \emph{{The Cosmological gravitational
				wave background from primordial density perturbations}},
		\href{https://doi.org/10.1103/PhysRevD.75.123518}{\emph{Phys. Rev. D}
			{\bfseries 75} (2007) 123518}
		[\href{https://arxiv.org/abs/gr-qc/0612013}{{\ttfamily gr-qc/0612013}}].
		
		\bibitem{Baumann:2007zm}
		D.~Baumann, P.J.~Steinhardt, K.~Takahashi and K.~Ichiki, \emph{{Gravitational
				Wave Spectrum Induced by Primordial Scalar Perturbations}},
		\href{https://doi.org/10.1103/PhysRevD.76.084019}{\emph{Phys. Rev. D}
			{\bfseries 76} (2007) 084019}
		[\href{https://arxiv.org/abs/hep-th/0703290}{{\ttfamily hep-th/0703290}}].
		
		\bibitem{Alabidi:2012ex}
		L.~Alabidi, K.~Kohri, M.~Sasaki and Y.~Sendouda, \emph{{Observable Spectra of
				Induced Gravitational Waves from Inflation}},
		\href{https://doi.org/10.1088/1475-7516/2012/09/017}{\emph{JCAP} {\bfseries
				09} (2012) 017} [\href{https://arxiv.org/abs/1203.4663}{{\ttfamily
				1203.4663}}].
		
		\bibitem{Nakama:2016gzw}
		T.~Nakama, J.~Silk and M.~Kamionkowski, \emph{{Stochastic gravitational waves
				associated with the formation of primordial black holes}},
		\href{https://doi.org/10.1103/PhysRevD.95.043511}{\emph{Phys. Rev. D}
			{\bfseries 95} (2017) 043511}
		[\href{https://arxiv.org/abs/1612.06264}{{\ttfamily 1612.06264}}].
		
		\bibitem{Kohri:2018awv}
		K.~Kohri and T.~Terada, \emph{{Semianalytic calculation of gravitational wave
				spectrum nonlinearly induced from primordial curvature perturbations}},
		\href{https://doi.org/10.1103/PhysRevD.97.123532}{\emph{Phys. Rev. D}
			{\bfseries 97} (2018) 123532}
		[\href{https://arxiv.org/abs/1804.08577}{{\ttfamily 1804.08577}}].
		
		\bibitem{Cheng:2018yyr}
		S.-L.~Cheng, W.~Lee and K.-W.~Ng, \emph{{Primordial black holes and associated
				gravitational waves in axion monodromy inflation}},
		\href{https://doi.org/10.1088/1475-7516/2018/07/001}{\emph{JCAP} {\bfseries
				07} (2018) 001} [\href{https://arxiv.org/abs/1801.09050}{{\ttfamily
				1801.09050}}].
		
		\bibitem{Cai:2019amo}
		R.-G.~Cai, S.~Pi, S.-J.~Wang and X.-Y.~Yang, \emph{{Resonant multiple peaks in
				the induced gravitational waves}},
		\href{https://doi.org/10.1088/1475-7516/2019/05/013}{\emph{JCAP} {\bfseries
				05} (2019) 013} [\href{https://arxiv.org/abs/1901.10152}{{\ttfamily
				1901.10152}}].
		
		\bibitem{Cai:2018dig}
		R.-g.~Cai, S.~Pi and M.~Sasaki, \emph{{Gravitational Waves Induced by
				non-Gaussian Scalar Perturbations}},
		\href{https://doi.org/10.1103/PhysRevLett.122.201101}{\emph{Phys. Rev. Lett.}
			{\bfseries 122} (2019) 201101}
		[\href{https://arxiv.org/abs/1810.11000}{{\ttfamily 1810.11000}}].
		
		\bibitem{Cai:2019elf}
		R.-G.~Cai, S.~Pi, S.-J.~Wang and X.-Y.~Yang, \emph{{Pulsar Timing Array
				Constraints on the Induced Gravitational Waves}},
		\href{https://doi.org/10.1088/1475-7516/2019/10/059}{\emph{JCAP} {\bfseries
				10} (2019) 059} [\href{https://arxiv.org/abs/1907.06372}{{\ttfamily
				1907.06372}}].
		
		\bibitem{Cai:2019bmk}
		R.-G.~Cai, Z.-K.~Guo, J.~Liu, L.~Liu and X.-Y.~Yang, \emph{{Primordial black
				holes and gravitational waves from parametric amplification of curvature
				perturbations}},
		\href{https://doi.org/10.1088/1475-7516/2020/06/013}{\emph{JCAP} {\bfseries
				06} (2020) 013} [\href{https://arxiv.org/abs/1912.10437}{{\ttfamily
				1912.10437}}].
		
		\bibitem{Cai:2020fnq}
		R.-G.~Cai, Y.-C.~Ding, X.-Y.~Yang and Y.-F.~Zhou, \emph{{Constraints on a mixed
				model of dark matter particles and primordial black holes from the galactic
				511 keV line}},
		\href{https://doi.org/10.1088/1475-7516/2021/03/057}{\emph{JCAP} {\bfseries
				03} (2021) 057} [\href{https://arxiv.org/abs/2007.11804}{{\ttfamily
				2007.11804}}].
		
		\bibitem{Pi:2020otn}
		S.~Pi and M.~Sasaki, \emph{{Gravitational Waves Induced by Scalar Perturbations
				with a Lognormal Peak}},
		\href{https://doi.org/10.1088/1475-7516/2020/09/037}{\emph{JCAP} {\bfseries
				09} (2020) 037} [\href{https://arxiv.org/abs/2005.12306}{{\ttfamily
				2005.12306}}].
		
		\bibitem{Domenech:2020kqm}
		G.~Dom\`enech, S.~Pi and M.~Sasaki, \emph{{Induced gravitational waves as a
				probe of thermal history of the universe}},
		\href{https://doi.org/10.1088/1475-7516/2020/08/017}{\emph{JCAP} {\bfseries
				08} (2020) 017} [\href{https://arxiv.org/abs/2005.12314}{{\ttfamily
				2005.12314}}].
		
		\bibitem{Liu:2021jnw}
		L.~Liu, X.-Y.~Yang, Z.-K.~Guo and R.-G.~Cai, \emph{{Testing primordial black
				hole and measuring the Hubble constant with multiband gravitational-wave
				observations}},
		\href{https://doi.org/10.1088/1475-7516/2023/01/006}{\emph{JCAP} {\bfseries
				01} (2023) 006} [\href{https://arxiv.org/abs/2112.05473}{{\ttfamily
				2112.05473}}].
		
		\bibitem{Papanikolaou:2021uhe}
		T.~Papanikolaou, C.~Tzerefos, S.~Basilakos and E.N.~Saridakis, \emph{{Scalar
				induced gravitational waves from primordial black hole Poisson fluctuations
				in f(R) gravity}},
		\href{https://doi.org/10.1088/1475-7516/2022/10/013}{\emph{JCAP} {\bfseries
				10} (2022) 013} [\href{https://arxiv.org/abs/2112.15059}{{\ttfamily
				2112.15059}}].
		
		\bibitem{Papanikolaou:2022hkg}
		T.~Papanikolaou, C.~Tzerefos, S.~Basilakos and E.N.~Saridakis, \emph{{No
				constraints for f(T) gravity from gravitational waves induced from primordial
				black hole fluctuations}},
		\href{https://doi.org/10.1140/epjc/s10052-022-11157-4}{\emph{Eur. Phys. J. C}
			{\bfseries 83} (2023) 31} [\href{https://arxiv.org/abs/2205.06094}{{\ttfamily
				2205.06094}}].
		
		\bibitem{Meng:2022low}
		D.-S.~Meng, C.~Yuan and Q.-G.~Huang, \emph{{Primordial black holes generated by
				the non-minimal spectator field}},
		\href{https://doi.org/10.1007/s11433-022-2095-5}{\emph{Sci. China Phys. Mech.
				Astron.} {\bfseries 66} (2023) 280411}
		[\href{https://arxiv.org/abs/2212.03577}{{\ttfamily 2212.03577}}].
		
		\bibitem{Danzmann:1997hm}
		K.~Danzmann, \emph{{LISA: An ESA cornerstone mission for a gravitational wave
				observatory}}, \href{https://doi.org/10.1088/0264-9381/14/6/002}{\emph{Class.
				Quant. Grav.} {\bfseries 14} (1997) 1399}.
		
		\bibitem{Audley:2017drz}
		{\scshape LISA} collaboration, \emph{{Laser Interferometer Space Antenna}},
		\href{https://arxiv.org/abs/1702.00786}{{\ttfamily 1702.00786}}.
		
		\bibitem{Hu:2017mde}
		W.-R.~Hu and Y.-L.~Wu, \emph{{The Taiji Program in Space for gravitational wave
				physics and the nature of gravity}},
		\href{https://doi.org/10.1093/nsr/nwx116}{\emph{Natl. Sci. Rev.} {\bfseries
				4} (2017) 685}.
		
		\bibitem{Luo:2015ght}
		{\scshape TianQin} collaboration, \emph{{TianQin: a space-borne gravitational
				wave detector}},
		\href{https://doi.org/10.1088/0264-9381/33/3/035010}{\emph{Class. Quant.
				Grav.} {\bfseries 33} (2016) 035010}
		[\href{https://arxiv.org/abs/1512.02076}{{\ttfamily 1512.02076}}].
		
		\bibitem{Gong:2021gvw}
		Y.~Gong, J.~Luo and B.~Wang, \emph{{Concepts and status of Chinese space
				gravitational wave detection projects}},
		\href{https://doi.org/10.1038/s41550-021-01480-3}{\emph{Nature Astron.}
			{\bfseries 5} (2021) 881} [\href{https://arxiv.org/abs/2109.07442}{{\ttfamily
				2109.07442}}].
		
		\bibitem{Kawamura:2011zz}
		S.~Kawamura et~al., \emph{{The Japanese space gravitational wave antenna:
				DECIGO}}, \href{https://doi.org/10.1088/0264-9381/28/9/094011}{\emph{Class.
				Quant. Grav.} {\bfseries 28} (2011) 094011}.
		
		\bibitem{Akrami:2018odb}
		{\scshape Planck} collaboration, \emph{{Planck 2018 results. X. Constraints on
				inflation}}, \href{https://doi.org/10.1051/0004-6361/201833887}{\emph{Astron.
				Astrophys.} {\bfseries 641} (2020) A10}
		[\href{https://arxiv.org/abs/1807.06211}{{\ttfamily 1807.06211}}].
		
		\bibitem{Martin:2012pe}
		J.~Martin, H.~Motohashi and T.~Suyama, \emph{{Ultra Slow-Roll Inflation and the
				non-Gaussianity Consistency Relation}},
		\href{https://doi.org/10.1103/PhysRevD.87.023514}{\emph{Phys. Rev. D}
			{\bfseries 87} (2013) 023514}
		[\href{https://arxiv.org/abs/1211.0083}{{\ttfamily 1211.0083}}].
		
		\bibitem{Motohashi:2014ppa}
		H.~Motohashi, A.A.~Starobinsky and J.~Yokoyama, \emph{{Inflation with a
				constant rate of roll}},
		\href{https://doi.org/10.1088/1475-7516/2015/09/018}{\emph{JCAP} {\bfseries
				09} (2015) 018} [\href{https://arxiv.org/abs/1411.5021}{{\ttfamily
				1411.5021}}].
		
		\bibitem{Yi:2017mxs}
		Z.~Yi and Y.~Gong, \emph{{On the constant-roll inflation}},
		\href{https://doi.org/10.1088/1475-7516/2018/03/052}{\emph{JCAP} {\bfseries
				03} (2018) 052} [\href{https://arxiv.org/abs/1712.07478}{{\ttfamily
				1712.07478}}].
		
		\bibitem{Garcia-Bellido:2017mdw}
		J.~Garcia-Bellido and E.~Ruiz~Morales, \emph{{Primordial black holes from
				single field models of inflation}},
		\href{https://doi.org/10.1016/j.dark.2017.09.007}{\emph{Phys. Dark Univ.}
			{\bfseries 18} (2017) 47} [\href{https://arxiv.org/abs/1702.03901}{{\ttfamily
				1702.03901}}].
		
		\bibitem{Germani:2017bcs}
		C.~Germani and T.~Prokopec, \emph{{On primordial black holes from an inflection
				point}}, \href{https://doi.org/10.1016/j.dark.2017.09.001}{\emph{Phys. Dark
				Univ.} {\bfseries 18} (2017) 6}
		[\href{https://arxiv.org/abs/1706.04226}{{\ttfamily 1706.04226}}].
		
		\bibitem{Motohashi:2017kbs}
		H.~Motohashi and W.~Hu, \emph{{Primordial Black Holes and Slow-Roll
				Violation}}, \href{https://doi.org/10.1103/PhysRevD.96.063503}{\emph{Phys.
				Rev. D} {\bfseries 96} (2017) 063503}
		[\href{https://arxiv.org/abs/1706.06784}{{\ttfamily 1706.06784}}].
		
		\bibitem{Ezquiaga:2017fvi}
		J.M.~Ezquiaga, J.~Garcia-Bellido and E.~Ruiz~Morales, \emph{{Primordial Black
				Hole production in Critical Higgs Inflation}},
		\href{https://doi.org/10.1016/j.physletb.2017.11.039}{\emph{Phys. Lett. B}
			{\bfseries 776} (2018) 345}
		[\href{https://arxiv.org/abs/1705.04861}{{\ttfamily 1705.04861}}].
		
		\bibitem{Gong:2017qlj}
		H.~Di and Y.~Gong, \emph{{Primordial black holes and second order gravitational
				waves from ultra-slow-roll inflation}},
		\href{https://doi.org/10.1088/1475-7516/2018/07/007}{\emph{JCAP} {\bfseries
				07} (2018) 007} [\href{https://arxiv.org/abs/1707.09578}{{\ttfamily
				1707.09578}}].
		
		\bibitem{Ballesteros:2018wlw}
		G.~Ballesteros, J.~Beltran~Jimenez and M.~Pieroni, \emph{{Black hole formation
				from a general quadratic action for inflationary primordial fluctuations}},
		\href{https://doi.org/10.1088/1475-7516/2019/06/016}{\emph{JCAP} {\bfseries
				06} (2019) 016} [\href{https://arxiv.org/abs/1811.03065}{{\ttfamily
				1811.03065}}].
		
		\bibitem{Dalianis:2018frf}
		I.~Dalianis, A.~Kehagias and G.~Tringas, \emph{{Primordial black holes from
				\ensuremath{\alpha}-attractors}},
		\href{https://doi.org/10.1088/1475-7516/2019/01/037}{\emph{JCAP} {\bfseries
				01} (2019) 037} [\href{https://arxiv.org/abs/1805.09483}{{\ttfamily
				1805.09483}}].
		
		\bibitem{Bezrukov:2017dyv}
		F.~Bezrukov, M.~Pauly and J.~Rubio, \emph{{On the robustness of the primordial
				power spectrum in renormalized Higgs inflation}},
		\href{https://doi.org/10.1088/1475-7516/2018/02/040}{\emph{JCAP} {\bfseries
				02} (2018) 040} [\href{https://arxiv.org/abs/1706.05007}{{\ttfamily
				1706.05007}}].
		
		\bibitem{Kannike:2017bxn}
		K.~Kannike, L.~Marzola, M.~Raidal and H.~Veerm\"ae, \emph{{Single Field Double
				Inflation and Primordial Black Holes}},
		\href{https://doi.org/10.1088/1475-7516/2017/09/020}{\emph{JCAP} {\bfseries
				09} (2017) 020} [\href{https://arxiv.org/abs/1705.06225}{{\ttfamily
				1705.06225}}].
		
		\bibitem{Lin:2020goi}
		J.~Lin, Q.~Gao, Y.~Gong, Y.~Lu, C.~Zhang and F.~Zhang, \emph{{Primordial black
				holes and secondary gravitational waves from $k$ and $G$ inflation}},
		\href{https://doi.org/10.1103/PhysRevD.101.103515}{\emph{Phys. Rev. D}
			{\bfseries 101} (2020) 103515}
		[\href{https://arxiv.org/abs/2001.05909}{{\ttfamily 2001.05909}}].
		
		\bibitem{Lin:2021vwc}
		J.~Lin, S.~Gao, Y.~Gong, Y.~Lu, Z.~Wang and F.~Zhang, \emph{{Primordial black
				holes and scalar induced gravitational waves from Higgs inflation with
				noncanonical kinetic term}},
		\href{https://doi.org/10.1103/PhysRevD.107.043517}{\emph{Phys. Rev. D}
			{\bfseries 107} (2023) 043517}
		[\href{https://arxiv.org/abs/2111.01362}{{\ttfamily 2111.01362}}].
		
		\bibitem{Gao:2019sbz}
		Q.~Gao, Y.~Gong and Z.~Yi, \emph{{On the constant-roll inflation with large and
				small $\eta_H$}},
		\href{https://doi.org/10.3390/universe5110215}{\emph{Universe} {\bfseries 5}
			(2019) 215} [\href{https://arxiv.org/abs/1901.04646}{{\ttfamily
				1901.04646}}].
		
		\bibitem{Gao:2020tsa}
		Q.~Gao, Y.~Gong and Z.~Yi, \emph{{Primordial black holes and secondary
				gravitational waves from natural inflation}},
		\href{https://doi.org/10.1016/j.nuclphysb.2021.115480}{\emph{Nucl. Phys. B}
			{\bfseries 969} (2021) 115480}
		[\href{https://arxiv.org/abs/2012.03856}{{\ttfamily 2012.03856}}].
		
		\bibitem{Gao:2021vxb}
		Q.~Gao, \emph{{Primordial black holes and secondary gravitational waves from
				chaotic inflation}},
		\href{https://doi.org/10.1007/s11433-021-1708-9}{\emph{Sci. China Phys. Mech.
				Astron.} {\bfseries 64} (2021) 280411}
		[\href{https://arxiv.org/abs/2102.07369}{{\ttfamily 2102.07369}}].
		
		\bibitem{Yi:2020kmq}
		Z.~Yi, Y.~Gong, B.~Wang and Z.-h.~Zhu, \emph{{Primordial black holes and
				secondary gravitational waves from the Higgs field}},
		\href{https://doi.org/10.1103/PhysRevD.103.063535}{\emph{Phys. Rev. D}
			{\bfseries 103} (2021) 063535}
		[\href{https://arxiv.org/abs/2007.09957}{{\ttfamily 2007.09957}}].
		
		\bibitem{Yi:2020cut}
		Z.~Yi, Q.~Gao, Y.~Gong and Z.-h.~Zhu, \emph{{Primordial black holes and
				scalar-induced secondary gravitational waves from inflationary models with a
				noncanonical kinetic term}},
		\href{https://doi.org/10.1103/PhysRevD.103.063534}{\emph{Phys. Rev. D}
			{\bfseries 103} (2021) 063534}
		[\href{https://arxiv.org/abs/2011.10606}{{\ttfamily 2011.10606}}].
		
		\bibitem{Yi:2021lxc}
		Z.~Yi and Z.-H.~Zhu, \emph{{NANOGrav signal and LIGO-Virgo primordial black
				holes from the Higgs field}},
		\href{https://doi.org/10.1088/1475-7516/2022/05/046}{\emph{JCAP} {\bfseries
				05} (2022) 046} [\href{https://arxiv.org/abs/2105.01943}{{\ttfamily
				2105.01943}}].
		
		\bibitem{Yi:2022anu}
		Z.~Yi, \emph{{Primordial black holes and scalar-induced gravitational waves
				from the generalized Brans-Dicke theory}},
		\href{https://doi.org/10.1088/1475-7516/2023/03/048}{\emph{JCAP} {\bfseries
				03} (2023) 048} [\href{https://arxiv.org/abs/2206.01039}{{\ttfamily
				2206.01039}}].
		
		\bibitem{Zhang:2020uek}
		F.~Zhang, Y.~Gong, J.~Lin, Y.~Lu and Z.~Yi, \emph{{Primordial non-Gaussianity
				from G-inflation}},
		\href{https://doi.org/10.1088/1475-7516/2021/04/045}{\emph{JCAP} {\bfseries
				04} (2021) 045} [\href{https://arxiv.org/abs/2012.06960}{{\ttfamily
				2012.06960}}].
		
		\bibitem{Pi:2017gih}
		S.~Pi, Y.-l.~Zhang, Q.-G.~Huang and M.~Sasaki, \emph{{Scalaron from
				$R^2$-gravity as a heavy field}},
		\href{https://doi.org/10.1088/1475-7516/2018/05/042}{\emph{JCAP} {\bfseries
				05} (2018) 042} [\href{https://arxiv.org/abs/1712.09896}{{\ttfamily
				1712.09896}}].
		
		\bibitem{Kamenshchik:2018sig}
		A.Y.~Kamenshchik, A.~Tronconi, T.~Vardanyan and G.~Venturi,
		\emph{{Non-Canonical Inflation and Primordial Black Holes Production}},
		\href{https://doi.org/10.1016/j.physletb.2019.02.036}{\emph{Phys. Lett. B}
			{\bfseries 791} (2019) 201}
		[\href{https://arxiv.org/abs/1812.02547}{{\ttfamily 1812.02547}}].
		
		\bibitem{Fu:2019ttf}
		C.~Fu, P.~Wu and H.~Yu, \emph{{Primordial Black Holes from Inflation with
				Nonminimal Derivative Coupling}},
		\href{https://doi.org/10.1103/PhysRevD.100.063532}{\emph{Phys. Rev. D}
			{\bfseries 100} (2019) 063532}
		[\href{https://arxiv.org/abs/1907.05042}{{\ttfamily 1907.05042}}].
		
		\bibitem{Fu:2019vqc}
		C.~Fu, P.~Wu and H.~Yu, \emph{{Scalar induced gravitational waves in inflation
				with gravitationally enhanced friction}},
		\href{https://doi.org/10.1103/PhysRevD.101.023529}{\emph{Phys. Rev. D}
			{\bfseries 101} (2020) 023529}
		[\href{https://arxiv.org/abs/1912.05927}{{\ttfamily 1912.05927}}].
		
		\bibitem{Dalianis:2019vit}
		I.~Dalianis, S.~Karydas and E.~Papantonopoulos, \emph{{Generalized Non-Minimal
				Derivative Coupling: Application to Inflation and Primordial Black Hole
				Production}},
		\href{https://doi.org/10.1088/1475-7516/2020/06/040}{\emph{JCAP} {\bfseries
				06} (2020) 040} [\href{https://arxiv.org/abs/1910.00622}{{\ttfamily
				1910.00622}}].
		
		\bibitem{Gundhi:2020zvb}
		A.~Gundhi and C.F.~Steinwachs, \emph{{Scalaron\textendash{}Higgs inflation
				reloaded: Higgs-dependent scalaron mass and primordial black hole dark
				matter}}, \href{https://doi.org/10.1140/epjc/s10052-021-09225-2}{\emph{Eur.
				Phys. J. C} {\bfseries 81} (2021) 460}
		[\href{https://arxiv.org/abs/2011.09485}{{\ttfamily 2011.09485}}].
		
		\bibitem{Cheong:2019vzl}
		D.Y.~Cheong, S.M.~Lee and S.C.~Park, \emph{{Primordial black holes in
				Higgs-$R^2$ inflation as the whole of dark matter}},
		\href{https://doi.org/10.1088/1475-7516/2021/01/032}{\emph{JCAP} {\bfseries
				01} (2021) 032} [\href{https://arxiv.org/abs/1912.12032}{{\ttfamily
				1912.12032}}].
		
		\bibitem{Zhang:2021rqs}
		F.~Zhang, \emph{{Primordial black holes and scalar induced gravitational waves
				from the E model with a Gauss-Bonnet term}},
		\href{https://doi.org/10.1103/PhysRevD.105.063539}{\emph{Phys. Rev. D}
			{\bfseries 105} (2022) 063539}
		[\href{https://arxiv.org/abs/2112.10516}{{\ttfamily 2112.10516}}].
		
		\bibitem{Zhang:2021vak}
		F.~Zhang, J.~Lin and Y.~Lu, \emph{{Double-peaked inflation model: Scalar
				induced gravitational waves and primordial-black-hole suppression from
				primordial non-Gaussianity}},
		\href{https://doi.org/10.1103/PhysRevD.104.063515}{\emph{Phys. Rev. D}
			{\bfseries 104} (2021) 063515}
		[\href{https://arxiv.org/abs/2106.10792}{{\ttfamily 2106.10792}}].
		
		\bibitem{Kawai:2021edk}
		S.~Kawai and J.~Kim, \emph{{Primordial black holes from Gauss-Bonnet-corrected
				single field inflation}},
		\href{https://doi.org/10.1103/PhysRevD.104.083545}{\emph{Phys. Rev. D}
			{\bfseries 104} (2021) 083545}
		[\href{https://arxiv.org/abs/2108.01340}{{\ttfamily 2108.01340}}].
		
		\bibitem{Cai:2021wzd}
		R.-G.~Cai, C.~Chen and C.~Fu, \emph{{Primordial black holes and stochastic
				gravitational wave background from inflation with a noncanonical spectator
				field}}, \href{https://doi.org/10.1103/PhysRevD.104.083537}{\emph{Phys. Rev.
				D} {\bfseries 104} (2021) 083537}
		[\href{https://arxiv.org/abs/2108.03422}{{\ttfamily 2108.03422}}].
		
		\bibitem{Chen:2021nio}
		P.~Chen, S.~Koh and G.~Tumurtushaa, \emph{{Primordial black holes and induced
				gravitational waves from inflation in the Horndeski theory of gravity}},
		\href{https://arxiv.org/abs/2107.08638}{{\ttfamily 2107.08638}}.
		
		\bibitem{Zheng:2021vda}
		R.~Zheng, J.~Shi and T.~Qiu, \emph{{On primordial black holes and secondary
				gravitational waves generated from inflation with solo/multi-bumpy potential
				*}}, \href{https://doi.org/10.1088/1674-1137/ac42bd}{\emph{Chin. Phys. C}
			{\bfseries 46} (2022) 045103}
		[\href{https://arxiv.org/abs/2106.04303}{{\ttfamily 2106.04303}}].
		
		\bibitem{Karam:2022nym}
		A.~Karam, N.~Koivunen, E.~Tomberg, V.~Vaskonen and H.~Veerm\"ae, \emph{{Anatomy
				of single-field inflationary models for primordial black holes}},
		\href{https://doi.org/10.1088/1475-7516/2023/03/013}{\emph{JCAP} {\bfseries
				03} (2023) 013} [\href{https://arxiv.org/abs/2205.13540}{{\ttfamily
				2205.13540}}].
		
		\bibitem{Ashoorioon:2019xqc}
		A.~Ashoorioon, A.~Rostami and J.T.~Firouzjaee, \emph{{EFT compatible PBHs:
				effective spawning of the seeds for primordial black holes during
				inflation}}, \href{https://doi.org/10.1007/JHEP07(2021)087}{\emph{JHEP}
			{\bfseries 07} (2021) 087}
		[\href{https://arxiv.org/abs/1912.13326}{{\ttfamily 1912.13326}}].
		
		\bibitem{Malik:2008im}
		K.A.~Malik and D.~Wands, \emph{{Cosmological perturbations}},
		\href{https://doi.org/10.1016/j.physrep.2009.03.001}{\emph{Phys. Rept.}
			{\bfseries 475} (2009) 1} [\href{https://arxiv.org/abs/0809.4944}{{\ttfamily
				0809.4944}}].
		
		\bibitem{Espinosa:2018eve}
		J.R.~Espinosa, D.~Racco and A.~Riotto, \emph{{A Cosmological Signature of the
				SM Higgs Instability: Gravitational Waves}},
		\href{https://doi.org/10.1088/1475-7516/2018/09/012}{\emph{JCAP} {\bfseries
				09} (2018) 012} [\href{https://arxiv.org/abs/1804.07732}{{\ttfamily
				1804.07732}}].
		
		\bibitem{Lu:2019sti}
		Y.~Lu, Y.~Gong, Z.~Yi and F.~Zhang, \emph{{Constraints on primordial curvature
				perturbations from primordial black hole dark matter and secondary
				gravitational waves}},
		\href{https://doi.org/10.1088/1475-7516/2019/12/031}{\emph{JCAP} {\bfseries
				12} (2019) 031} [\href{https://arxiv.org/abs/1907.11896}{{\ttfamily
				1907.11896}}].
		
		\bibitem{Ando:2017veq}
		K.~Ando, K.~Inomata, M.~Kawasaki, K.~Mukaida and T.T.~Yanagida,
		\emph{{Primordial black holes for the LIGO events in the axionlike curvaton
				model}}, \href{https://doi.org/10.1103/PhysRevD.97.123512}{\emph{Phys. Rev.
				D} {\bfseries 97} (2018) 123512}
		[\href{https://arxiv.org/abs/1711.08956}{{\ttfamily 1711.08956}}].
		
		\bibitem{Vaskonen:2020lbd}
		V.~Vaskonen and H.~Veerm\"ae, \emph{{Did NANOGrav see a signal from primordial
				black hole formation?}},
		\href{https://doi.org/10.1103/PhysRevLett.126.051303}{\emph{Phys. Rev. Lett.}
			{\bfseries 126} (2021) 051303}
		[\href{https://arxiv.org/abs/2009.07832}{{\ttfamily 2009.07832}}].
		
		\bibitem{DeLuca:2020agl}
		V.~De~Luca, G.~Franciolini and A.~Riotto, \emph{{NANOGrav Data Hints at
				Primordial Black Holes as Dark Matter}},
		\href{https://doi.org/10.1103/PhysRevLett.126.041303}{\emph{Phys. Rev. Lett.}
			{\bfseries 126} (2021) 041303}
		[\href{https://arxiv.org/abs/2009.08268}{{\ttfamily 2009.08268}}].
		
		\bibitem{Planck:2018jri}
		{\scshape Planck} collaboration, \emph{{Planck 2018 results. X. Constraints on
				inflation}}, \href{https://doi.org/10.1051/0004-6361/201833887}{\emph{Astron.
				Astrophys.} {\bfseries 641} (2020) A10}
		[\href{https://arxiv.org/abs/1807.06211}{{\ttfamily 1807.06211}}].
		
		\bibitem{Ashton:2018jfp}
		G.~Ashton et~al., \emph{{BILBY: A user-friendly Bayesian inference library for
				gravitational-wave astronomy}},
		\href{https://doi.org/10.3847/1538-4365/ab06fc}{\emph{Astrophys. J. Suppl.}
			{\bfseries 241} (2019) 27}
		[\href{https://arxiv.org/abs/1811.02042}{{\ttfamily 1811.02042}}].
		
		\bibitem{NestedSampling}
		J.~Skilling, \emph{Nested sampling}, {\emph{AIP Conf. Proc.} (2004) 395}.
		
		\bibitem{Moore:2021ibq}
		C.J.~Moore and A.~Vecchio, \emph{{Ultra-low-frequency gravitational waves from
				cosmological and astrophysical processes}},
		\href{https://doi.org/10.1038/s41550-021-01489-8}{\emph{Nature Astron.}
			{\bfseries 5} (2021) 1268}
		[\href{https://arxiv.org/abs/2104.15130}{{\ttfamily 2104.15130}}].
		
		\bibitem{Lamb:2023jls}
		W.G.~Lamb, S.R.~Taylor and R.~van Haasteren, \emph{{The Need For Speed: Rapid
				Refitting Techniques for Bayesian Spectral Characterization of the
				Gravitational Wave Background Using PTAs}},
		\href{https://arxiv.org/abs/2303.15442}{{\ttfamily 2303.15442}}.
		
		\bibitem{EPTA:2023hof}
		{\scshape EPTA} collaboration, \emph{{Practical approaches to analyzing PTA
				data: Cosmic strings with six pulsars}},
		\href{https://arxiv.org/abs/2306.12234}{{\ttfamily 2306.12234}}.
		
		\bibitem{Jeong:2014gna}
		D.~Jeong, J.~Pradler, J.~Chluba and M.~Kamionkowski, \emph{{Silk damping at a
				redshift of a billion: a new limit on small-scale adiabatic perturbations}},
		\href{https://doi.org/10.1103/PhysRevLett.113.061301}{\emph{Phys. Rev. Lett.}
			{\bfseries 113} (2014) 061301}
		[\href{https://arxiv.org/abs/1403.3697}{{\ttfamily 1403.3697}}].
		
		\bibitem{Inomata:2016uip}
		K.~Inomata, M.~Kawasaki and Y.~Tada, \emph{{Revisiting constraints on small
				scale perturbations from big-bang nucleosynthesis}},
		\href{https://doi.org/10.1103/PhysRevD.94.043527}{\emph{Phys. Rev. D}
			{\bfseries 94} (2016) 043527}
		[\href{https://arxiv.org/abs/1605.04646}{{\ttfamily 1605.04646}}].
		
		\bibitem{Fixsen:1996nj}
		D.J.~Fixsen, E.S.~Cheng, J.M.~Gales, J.C.~Mather, R.A.~Shafer and E.L.~Wright,
		\emph{{The Cosmic Microwave Background spectrum from the full COBE FIRAS data
				set}}, \href{https://doi.org/10.1086/178173}{\emph{Astrophys. J.} {\bfseries
				473} (1996) 576} [\href{https://arxiv.org/abs/astro-ph/9605054}{{\ttfamily
				astro-ph/9605054}}].
		
		\bibitem{Chluba:2012we}
		J.~Chluba, A.L.~Erickcek and I.~Ben-Dayan, \emph{{Probing the inflaton:
				Small-scale power spectrum constraints from measurements of the CMB energy
				spectrum}},
		\href{https://doi.org/10.1088/0004-637X/758/2/76}{\emph{Astrophys. J.}
			{\bfseries 758} (2012) 76} [\href{https://arxiv.org/abs/1203.2681}{{\ttfamily
				1203.2681}}].
		
	\end{thebibliography}
	
	\providecommand{\href}[2]{#2}\begingroup\raggedright\endgroup
	
\end{document}